\documentclass[]{aa}
\usepackage[varg]{txfonts}
\usepackage{natbib}
\usepackage[final]{hyperref}
\usepackage{xcolor}
\usepackage{multirow}
\usepackage[switch]{lineno}
\usepackage{booktabs}
\usepackage{siunitx}
\usepackage{upgreek}

\begin{document}

\title{Galaxy and Mass Assembly (GAMA)}
\subtitle{Tracing galaxy environment using the marked correlation function}

\author{
        U.~Sureshkumar \inst{\ref{aff:oauj}} 
        \and A.~Durkalec \inst{\ref{aff:ncbj}} 
        \and A.~Pollo \inst{\ref{aff:oauj},\ref{aff:ncbj}} 
        \and M.~Bilicki \inst{\ref{aff:ctp_warsaw}} 
        \and J.~Loveday \inst{\ref{aff:loveday}}
        \and D.~J.~Farrow \inst{\ref{aff:farrow}}
        \and B.~W.~Holwerda \inst{\ref{aff:holwerda}}
        \and A.~M.~Hopkins \inst{\ref{aff:hopkins}}
        \and J.~Liske \inst{\ref{aff:liske}}
        \and K.~A.~Pimbblet \inst{\ref{aff:pimbblet}}
        \and E.~N.~Taylor \inst{\ref{aff:taylor}}
        \and A.~H.~Wright \inst{\ref{aff:wright}}
        }

\institute{
    Astronomical Observatory of the Jagiellonian University, ul. Orla 171, 30-244 Krak\'{o}w, Poland \\\email{usureshkumar@oa.uj.edu.pl} 
    \label{aff:oauj}
    \and
    National Centre for Nuclear Research, ul. Pasteura 7, 02-093 Warsaw, Poland 
    \label{aff:ncbj}
    \and
    Center for Theoretical Physics, Polish Academy of Sciences, al. Lotnik\'{o}w 32/46, 02-668 Warsaw, Poland 
    \label{aff:ctp_warsaw}
    \and
    Astronomy Centre, University of Sussex, Falmer, Brighton BN1 9QH, UK
    \label{aff:loveday}
    \and
    Max-Planck-Institut für extraterrestrische Physik, Giessenbachstrasse 1, 85748 Garching, Germany 
    \label{aff:farrow}
    \and
    Department of Physics and Astronomy, 102 Natural Science Building, University of Louisville, Louisville KY 40292, USA
    \label{aff:holwerda}
    \and
    Australian Astronomical Optics, Macquarie University, 105 Delhi Rd, North Ryde, NSW 2113, Australia
    \label{aff:hopkins}
    \and
    Hamburger Sternwarte, Universität Hamburg, Gojenbergsweg 112, 21029 Hamburg, Germany
    \label{aff:liske}
    \and
    E.~A.~Milne Centre for Astrophysics, University of Hull, Cottingham Road, Kingston-upon-Hull, HU6 7RX, UK
    \label{aff:pimbblet}
    \and
    Centre for Astrophysics and Supercomputing, Swinburne University of Technology, Hawthorn 3122, Australia
    \label{aff:taylor}
    \and
    Ruhr University Bochum, Faculty of Physics and Astronomy, Astronomical Institute (AIRUB), German Centre for Cosmological Lensing, 44780 Bochum, Germany
    \label{aff:wright}
    }

\date{Received 5 February 2021 / Accepted 1 June 2021}

\abstract
%Context
{
Galaxies are biased tracers of the underlying network of dark matter.
The strength of this bias depends on various galaxy properties and on redshift.
One of the methods used to study these dependences of the bias is measurement of galaxy clustering.
Such studies are made using galaxy samples from various catalogues, which frequently bear their own problems related to sample selection methods.
It is therefore crucial to understand how sample choice influences clustering measurements and which galaxy property is the most direct tracer of the galaxy environment.
}
%Aim
{
We investigate how different galaxy properties, such as luminosities in the $u, g, r, J,$ and $K$ bands, stellar mass, star formation rate, and specific star formation rate, trace the environment in the local universe.
We also study the effect of survey flux limits on galaxy clustering measurements.
}
%Method
{
We measure the two-point correlation function (2pCF) and marked correlation functions (MCFs) using the aforementioned properties as marks. 
We use a nearly stellar-mass-complete galaxy sample in the redshift range $0.1 < z < 0.16$ from the Galaxy and Mass Assembly (GAMA) survey with a flux limit of $r < 19.8$.
Further, we impose a brighter flux limit of $r < 17.8$ on our sample and repeated the measurements to study how this affects galaxy clustering analysis.
We compare our results to measurements from the Sloan Digital Sky Survey (SDSS) with flux limits of $r < 17.8$ and $r < 16.8$.
}
%Result
{
We show that the stellar mass is the most direct tracer of galaxy environment, the $K$-band luminosity being a good substitute, although such a proxy sample misses close pairs of evolved, red galaxies.
We also show that the $u$-band luminosity can be a proxy to the star formation rate in the context of galaxy clustering.
We observe an effect of the survey flux limit on clustering studies; samples with a higher flux limit (smaller magnitude) miss some information about close pairs of starburst galaxies.
}
%Conclusion
{}
 

\titlerunning{Tracing galaxy environment using the marked correlation function in GAMA}
\authorrunning{U. Sureshkumar et al.}

\maketitle

\section{Introduction}\label{sec:introduction}

Local galaxy observations reveal the large-scale structure (LSS) of the Universe to be a rich network of filaments, walls, nodes, and voids \citep{delapparent1986, bond1996, alpaslan2014}.
According to the $\Lambda$ cold dark matter ($\Lambda$CDM) cosmological model these structures are built of two main elements: baryonic and dark matter. 
The former exists in the form of stars, gas, and dust; these can be traced at different wavelengths using large sky surveys. 
Dark matter, however, cannot be observed directly, although it is gravitationally dominant. 
Therefore, we rely on visible baryonic matter observations to indirectly trace the underlying dark matter distribution.

One of the methods used to connect baryonic and dark matter involves measurements of the galaxy two-point correlation function \citep[2pCF;][]{peebles1980}. 
This powerful statistical tool describes the spatial distribution of galaxies and has been extensively used in the past to quantify their clustering and its various dependences: on luminosity \citep{norberg2001, pollo2006, torre2007, meneux2009, zehavi2011, marulli2013, guo2015, farrow2015_gama_cf}, stellar mass \citep{meneux2008, marulli2013, beutler2013, dolley2014, skibba2015, durkalec2018}, star formation rate \citep[SFR;][]{hartley2010, mostek2013}, colour \citep{zehavi2005, coil2008, skibba2014_combining_fields}, and spectral type \citep{norberg2002, meneux2006}.
The general conclusion from all these studies is that galaxy clustering strongly depends on galaxy properties. 
More luminous, massive, redder, and early-type galaxies exhibit stronger clustering, and tend to exist in denser regions of the universe, than their less massive, bluer, and later-type counterparts.

These observations can be explained in the framework of the $\Lambda$CDM cosmology and hierarchical model of structure formation.
Small density fluctuations in the early universe evolved under gravity to form the present LSS \citep{springer2005}. 
The initially stronger over-densities evolved faster, thus resulting in the formation of self-bound clumps of dark matter called dark matter haloes.
Such haloes provided the gravitational potential to trap the baryonic matter and thereby form galaxies at their centres \citep{press_schechter_1974, white&rees1978}.
Therefore, it is expected that the properties of the parent halo play a major role in defining galaxy properties such as luminosity, stellar mass, colour, and SFR. 
The halo mass is believed to greatly influence the halo clustering and hence the properties of hosted galaxies \citep[e.g.][]{zheng2005, more2009, gu2016}; this is frequently referred to as `halo bias'.
Large and massive haloes have potentials strong enough to form bigger, more massive, and more luminous galaxies. 
However, it has also been shown that the clustering of haloes has a dependence on properties other than halo mass---in large part the halo assembly history that is commonly referred to as `halo assembly bias' \citep{zentner2014, croton2007, mao2018}.
This means that halo properties correlate with environment \citep{sheth&tormen2004}.
These two dependences, that is those between halo properties and galaxy properties, and between halo properties and environment, prompt a correlation between galaxy properties and environment  \citep[see][for a review]{wechsler2018}.

Studies of these dependences between galaxy properties and environment are crucial to understand structure formation in the universe, and over the past decade there has been remarkable progress in the development of galaxy formation models describing connections between dark matter haloes and their galaxies \citep[see][for a review]{somerville2015}. 
Methods used in these models include numerical hydrodynamic techniques \citep[e.g.][]{mccarthy2012, vogelsberger2013, kannan2014}, semi-analytic models \citep[SAM; e.g.][]{baugh2006, benson2012, linke2020}, and even empirical methods in which physical constraints are taken entirely from observations \citep[e.g.][]{yang2012, moster2020, grylls2020}.
We are now able to simulate the physics of galaxy formation and, to some extent, link galaxy properties to the host halo properties.
For example, simulations have been used, such as {\small UNIVERSEMACHINE} \citep{behroozi2019}, which reasonably parametrises the galaxy growth-halo assembly correlation, and {\small SHARK} \citep{lagos2018}, which agrees fairly well with observations \citep{bravo2020}.
But there is still a need for improvement in understanding the process of galaxy formation \citep[see][for a review]{naab2017}.
There has not yet been a method that would perfectly reconstruct the observed dependence of galaxy properties on halo properties and environment and the dependences of galaxy clustering on galaxy properties.
Hence, a better understanding of how different galaxy properties trace the environment is needed to establish better constraints on galaxy formation and evolution models. 
It would be preferred if this understanding came from galaxy observations rather than simulations.

There is a problem that has to be faced here: galaxy clustering strength depends on the photometric passband in which the galaxies are selected for measurement \citep{milliard2007, torre2007, zehavi2011, skibba2014_combining_fields}. 
In other words, it is common that different works report various clustering strengths for galaxies selected using different methods. 
For example, in the optical range, \citet{zehavi2005} measured correlation functions (CFs) in volume-limited samples of galaxies from the Sloan Digital Sky Survey (SDSS) selected in bins of the $r$-band absolute magnitude.
These authors observed that galaxies that are brighter in the $r$ band show stronger clustering than the fainter galaxies.
Similar behaviour of clustering was observed for the $B$ band from the VIMOS survey \citep{marulli2013}, $g$ band from PRIMUS \citep{skibba2014_combining_fields}, and $K$ band from HiZELS \citep{sobral2010}.
Measurements in the $u$ band, however, show an opposite trend. 
Galaxies luminous in that band tend to exist in low-density regions of the universe, whereas their $u$-band fainter counterparts are preferentially found in high-density locations \citep{deng2012}.
Additionally, measurements based on farther infrared (IR) indicate stronger clustering than that measured for galaxies observed at optical wavelengths \citep{oliver2004, pollo2013_akarinorth, pollo2013_akarifull}.
Moreover, \citet{heinis2004} and \citet{milliard2007} reported  weaker clustering of ultra-violet (UV) galaxies in the local universe compared to optical and IR galaxies.
All these results imply that galaxies selected based on different properties trace the local environment differently.

The main aim of this paper is to demonstrate how different galaxy properties trace the small-scale galaxy clustering.
In particular, we show which property can be a better tracer of galaxy environment.
Environment has been defined in many different ways in the past \citep{muldrew2012}; in this work we define it as the galaxy over-density around the object.

We also study the possible influence of selection methods on clustering results.
In particular we show which photometric passband best serves as a proxy for stellar mass in the absence thereof, and how survey flux limitations can influence clustering results.   
Frequently, in the literature, luminosity and stellar mass are considered to be one-to-one correlated: more luminous galaxies are assumed to be more massive \citep{blanton&moustakas2009}.
In particular, near-infrared (NIR) luminosity is known to be a good proxy of the galaxy stellar mass \citep{kochanek2001}.
Mid-infrared (MIR) fluxes, particularly those with 3.4 $\mu$m and 4.6 $\mu$m wavelengths, are also reliable tracers of galaxy stellar mass \citep{jarrett2013, cluver2014}.
For instance, $K$-band selected samples are used to construct stellar mass limited samples with high completeness \citep{vandokkum2006, taylor2009}.
However, it is yet unclear if such a sample can be a perfect proxy for clustering measurements.
Better understanding on this issue gives us a better idea on the cautions to be taken while these kinds of proxy samples are used for clustering studies.

In addition, galaxy surveys are inevitably flux-limited.
Therefore, extra care has to be taken while working with stellar mass selected samples extracted from such surveys.
They tend to miss galaxies that are massive enough to pass the mass selection, but not luminous enough to reach the flux limit of the survey \citep{meneux2008, meneux2009, marulli2013}. 
This effect makes such samples incomplete and hence can influence the galaxy clustering measurements. 
In this work we try to understand how the flux limit of a survey affects the measured clustering and what steps are needed to account for resulting inaccuracies.

In our work we use the Galaxy and Mass Assembly spectroscopic survey \citep[GAMA;][]{driver2009_gama_gen}. 
We choose GAMA over SDSS owing to its high completeness ($> 98.5\%$) down to $r_\text{petro} < 19.8$ (2 mag fainter than SDSS).
Moreover, GAMA does not suffer from fibre collisions that affect the close galaxy pairs in SDSS \citep{robotham2010_gama_tiling}.
It also provides reliable measurements of absolute magnitudes in a wide wavelength range, stellar masses, and SFR.
The GAMA survey has been used for various aspects of environmental effects, in particular the impact of group, cluster, local, and large-scale environment on galaxy properties \citep{wijesinghe2012, burton2013, brough2013, mcnaught2014, alpaslan2015, davies2016, schaefer2017, grootes2017, barsanti2018, wang2018, davies2019_1, davies2019_2, schaefer2019, vazquez2020}.

Our methods rely on marked statistics tools \citep{stoyan&stoyan1994}, in particular on measurements of the galaxy marked correlation function (MCF), which has been proven to be very sensitive to the environment \citep{sheth&tormen2004, skibba2013}. 
In this method, each galaxy is assigned a `mark' that is defined as any measurable property of the galaxy. 
The MCF accounts for the clustering of positions of the marks (i.e. galaxies of a given property).
Hence, MCF measurements with different galaxy properties as marks help us to study how these different properties trace the galaxy clustering, particularly on small scales \citep{sheth2005}.
Marked statistics has been widely used to show that closer pairs of galaxies are more luminous, redder, and older than pairs that have larger separations \citep{beisbart2000, skibba2006, sheth2006}.
Additionally, \citet{sheth2005_galform_models} showed that these observations are in qualitative agreement with the semi-analytic galaxy formation models.
\citet{gunawardhana2018} used marked statistics on a set of luminosity- and stellar-mass-selected galaxy samples from GAMA with SFR, specific SFR (sSFR), and $(g-r)_\text{rest}$ colour as marks.
They observed that sSFR is a better tracer of interactions between star-forming galaxies than colour.
\citet{riggs2021} uses MCF to explore the clustering of galaxy groups in GAMA.

In our study we compute the MCFs on stellar-mass-selected samples in the redshift range $0.1 < z < 0.16$ using absolute magnitudes in the $u, g, r, J,$ and $K$ bands, stellar mass, SFR, and sSFR (SFR per unit stellar mass) of galaxies as marks.
Additionally, we explore the effect of apparent flux limits on the correlation between small-scale clustering and galaxy properties.
For this purpose, we impose various flux limits to the parent sample from the GAMA survey.
Further we compare the MCFs using different marks to see how these functions are affected by the change in flux limit.
We also compare the measurements in our GAMA sample with those from SDSS.
We select samples from SDSS because it provides a larger number of galaxies to brighter flux limits than GAMA.

This paper is structured as follows. 
In Sect.~\ref{sec:data} we describe properties of the GAMA survey and our sample selection method. 
Different clustering techniques and their definitions are described in Sect.~\ref{sec:measurement}. 
Clustering measurements are presented in Sect.~\ref{sec:results}. 
The results are discussed and compared with other works in Sect.~\ref{sec:discussion} and finally concluded in Sect.~\ref{sec:conclusion}. 

Throughout the paper, a flat $\Lambda$CDM cosmological model with $\Omega_{\text{M}}=0.3$ and $\Omega_\Lambda=0.7$ is adopted and the Hubble constant is parametrised via $h=H_0/100 \rm \, km \, s^{-1} \, Mpc^{-1}$. 
All galaxy properties are measured using $h=0.7$.
The distances are expressed in comoving coordinates and are in units of $h^{-1} \mathrm{Mpc}$.

\section{Data}\label{sec:data}

\subsection{Galaxy and Mass Assembly}\label{sec:data_gama}

The GAMA survey is a spectroscopic and multiwavelength galaxy survey that aims to test the $\Lambda$CDM model of structure formation and to study the galaxy evolution through the latest generation of ground-based and space-borne, wide-field survey facilities \citep{driver2011_gama_coredata, liske2015_gama_dr2}. 
It provides a comprehensive survey of galaxy populations by bringing together data from eight ground-based surveys and four space missions. 
The GAMA covers three equatorial regions named G09, G12, and G15 and two southern regions G02 and G23. 
Detailed descriptions of the GAMA survey are provided in \citet{driver2009_gama_gen}, \citet{robotham2010_gama_tiling}, \citet{driver2011_gama_coredata}, and \citet{liske2015_gama_dr2}; below we briefly describe the survey details important in the context of our work.

We exploit the main $r$-band limited data from GAMA II equatorial regions with targets drawn primarily from SDSS DR7 \citep{abazajian2009_sdss_dr7}. 
For  extinction-corrected $r$-band Petrosian magnitudes \citep{petrosian1976} limited at $r_{\text{petro}} < 19.8$, GAMA provides high spatial completeness and an overall redshift completeness of $98.48\%$ in the equatorial regions. 
This excellent completeness of GAMA is achieved by repeated surveying of the same field \citep{robotham2010_gama_tiling}, thereby making GAMA an ideal survey for clustering measurements.  

In this study we select galaxies from the GAMA II main survey (\texttt{SURVEY\_CLASS} $\geq$ 4) in the equatorial regions with spectroscopic redshifts in the range $0.1 < z < 0.16$. 
The redshifts of GAMA objects were measured using the software \textsc{autoz}, as described in \citet{baldry2014}, and are corrected for the local flow via the \citet{tonry2010} model, tapered smoothly to the cosmic microwave background rest frame for $z \geq 0.03$. 
\citet{liske2015_gama_dr2} provide a detailed assessment of the quality and reliability of these redshifts.
We only use secure redshifts with quality flag \texttt{nQ} $\geq$  3, which assures that the redshift has $>90\%$ chance of being correct.
In addition to this redshift quality cut, we only select objects with \texttt{VIS\_CLASS} = 0, \texttt{VIS\_CLASS} = 1, or \texttt{VIS\_CLASS} = 255, so that we avoid sources that are visually classified to be deblends of star or parts of other galaxies \citep{baldry2010}.

As GAMA combines photometric data from several surveys, it has a very wide wavelength range, from X-ray to radio.
In this work, we make use of \textsc{StellarMassesLambdarv20} DMU \citep{taylor2011_gama_stellar_mass,wright2016}.
The stellar masses are based on the methods of \citet{taylor2011_gama_stellar_mass} applied to the \textsc{lambdar} photometry of \citet{wright2016}.
These methods are based on the stellar population synthesis (SPS) modelling of broadband photometry using stellar evolution models by \citet{bruzual2003}, assuming a \citet{chabrier2003} initial mass function and \citet{calzetti2000} dust law.
The absolute magnitudes are inferred from the SPS fits and are corrected for internal dust extinction and \textit{k}-corrected to $z=0$.
As the fits are constrained to the rest-frame wavelength range 300 - 1100 nm, the $J$ and $K$-band absolute magnitudes we use are extrapolations of the fit to data. 
The stellar masses and absolute magnitudes are \texttt{fluxscale} corrected as described in \citet{taylor2011_gama_stellar_mass} to account for the difference in aperture matched and S\'{e}rsic photometry.
Galaxies with physically unrealistic \texttt{fluxscale} values are not considered for our analysis. 
The SFRs and sSFRs are taken from the DMU \textsc{MagPhysv06} and are estimated using the energy balance spectral energy distribution (SED)-fitting code \textsc{magphys} \citep{dacunha2008_gama_sfr}.
All the quantities are derived for the concordance $(\Omega_{\text{M}}, \Omega_{\Lambda}, h) = (0.3, 0.7, 0.7)$ cosmology.
For clustering measurements, we use the GAMA random galaxy catalogue (\textsc{Randomsv02} DMU) by \citet{farrow2015_gama_cf}.
In the random catalogue, we assign stellar mass to each random galaxy by matching on the \texttt{CATAID} of the real galaxy.

\begin{table}
\caption{Properties of the GAMA equatorial regions used in this work.}
\begin{center}
\begin{tabular}{c c c c}
\toprule
\toprule
  \multicolumn{1}{c}{GAMA region} &
  \multicolumn{1}{c}{$N_\mathrm{gal}$} &
  \multicolumn{1}{c}{$z_\mathrm{median}$} &
  \multicolumn{1}{c}{Area [deg\textsuperscript{2}]} \\
\midrule
  G09 & 7863 & 0.14 & 60  \\
  G12 & 12652 & 0.13 & 60  \\
  G15 & 13940 & 0.13 & 60  \\
\midrule
  Total & 34455 & 0.13 & 180 \\
\bottomrule
\end{tabular}
\end{center}
\label{table:gama_regions}
\end{table}

\subsection{Sample selection}\label{sec:data_sampleselection}

Our full sample counts 34455 galaxies in the redshift range $0.1 < z < 0.16$ with apparent flux limit $r_\text{petro} < 19.8$ distributed over the three equatorial regions in the sky, for which the details such as the number of galaxies, median redshift, and total area are given in Table~\ref{table:gama_regions}. 
The aforementioned redshift range is chosen to optimally select volume-limited samples that include low-mass galaxies.
All these galaxies have reliable spectroscopic redshift and well-measured absolute magnitudes, stellar mass, SFR, and sSFR. 

\begin{figure}[h]
    \includegraphics[width=\linewidth]{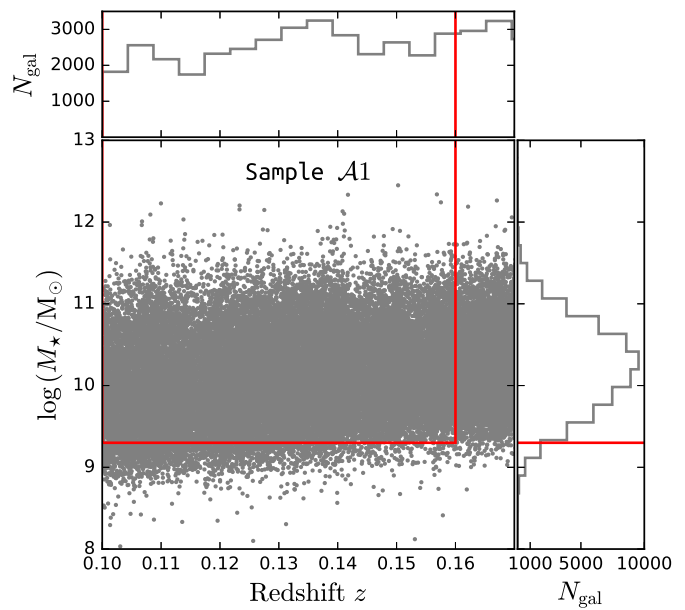}
    \caption{Selection of galaxy sample $\mathcal{A}1$ used in this work. 
    The grey dots represent the GAMA galaxies with flux limit $r<19.8$.
    The top and right histograms show the distribution of redshift and stellar mass, respectively.
    The red lines represent the stellar mass cut and the redshift limit of the sample $\mathcal{A}1$.}
    \label{fig:data_a1}
\end{figure}

To study the environmental dependence of luminosity in different bands, stellar mass, SFR, and sSFR, we define a nearly stellar-mass-complete sample by applying an additional stellar mass cut of $\log \, (M_{\star}/\mathrm{M_{\sun}})^\mathrm{min} = 9.3$. This is referred to as Sample $\mathcal{A}1$.
As a representative example of the selection technique, in Fig.~\ref{fig:data_a1} we show the selection cut for the sample $\mathcal{A}1$ with flux limit $r<19.8$.

\begin{figure}[h]
    \centering
    \includegraphics[width=\linewidth]{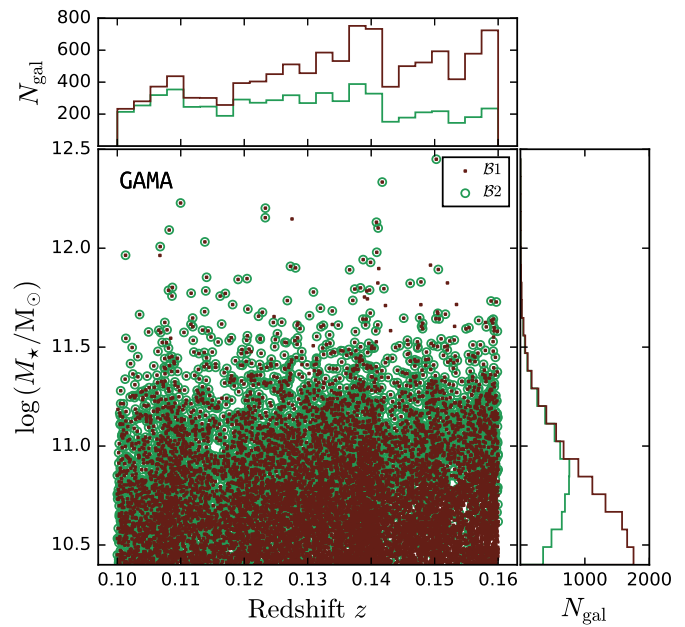}
    \caption{Redshift and stellar mass distribution of galaxies in stellar-mass-selected samples $\mathcal{B}1$ ($r < 19.8$; brown dots) and $\mathcal{B}2$ ($r < 17.8$; green circles).
    The top and right histograms show the distribution of redshift and stellar mass, respectively.}
    \label{fig:data_b1_b2}
\end{figure}

To further investigate how the environmental dependence of the galaxy properties varies with different flux limits, we select two stellar-mass-selected samples in the same redshift range with a stellar mass cut of $\log \, (M_{\star}/\mathrm{M_{\sun}})^\mathrm{min} = 10.4$ but different flux limits: Sample $\mathcal{B}1$ with $r<19.8$ and $\mathcal{B}2$ with $r<17.8$.
The stellar mass and redshift distribution of these samples are shown in Fig.~\ref{fig:data_b1_b2}.
The definition and properties of all the selected samples are given in Table~\ref{table:subsamples_definitions} and Table~\ref{table:subsamples_properties} respectively.
All the selected samples contain a sufficient number of galaxies for reliable clustering measurements. 
For all the selected samples, random samples are selected from \textsc{Randomsv02} DMU \citep{farrow2015_gama_cf} after applying corresponding $r$-band apparent magnitude cut and stellar mass cut. 

\begin{table}
\caption{Definitions of the galaxy samples in the redshift range $0.1 < z < 0.16$, as used in this study. }
\begin{center}
\begin{tabular}{c c c c c}
\toprule
\toprule
  \multicolumn{1}{c}{$\log \, \left( \frac{M_{\star}}{\mathrm{M_{\sun}}} \right)^{\mathrm{min}}$} &
  \multicolumn{1}{c}{Sample} &
  \multicolumn{1}{c}{Survey} &
  \multicolumn{1}{c}{$r_\mathrm{lim}$} &
  \multicolumn{1}{c}{$N_\mathrm{gal}$}
  \\ 
\midrule 
  9.3 & $\mathcal{A}1$ & GAMA & 19.8 & 32401 \\
  \midrule
  \multirow{2}{*}{10.4} & $\mathcal{B}1 $ & GAMA & 19.8 & 10706 \\
   & $ \mathcal{B}2 $ & GAMA & 17.8 & 5907 \\
  \midrule
  \multirow{5}{*}{10.8} & $\mathcal{C}1$ & GAMA & 19.8 & 3811 \\
  & $\mathcal{C}2$ & GAMA & 18.8 & 3752 \\
  & $\mathcal{C}3$ & GAMA & 17.8 & 3367 \\
  & $\mathcal{C}4$ & SDSS & 17.8 & 22772 \\
  & $\mathcal{C}5$ & SDSS & 16.8 & 11346 \\
\bottomrule
\end{tabular}
\end{center}
\tablefoot{The columns represent the stellar mass cut, sample label, survey of origin, flux limit, and number of galaxies.}
\label{table:subsamples_definitions}
\end{table}

\begin{table*}
\caption{Properties of the galaxy samples defined in Table~\ref{table:subsamples_definitions}.}
\begin{center}
\resizebox{\textwidth}{!}{
\begin{tabular}{c c c c c c c c c}
\toprule
\toprule
  \multicolumn{1}{c}{Sample} &
  \multicolumn{1}{c}{$M_u^{\mathrm{mean}\pm1\sigma}$} &
  \multicolumn{1}{c}{$M_g^{\mathrm{mean}\pm1\sigma}$} &
  \multicolumn{1}{c}{$M_r^{\mathrm{mean}\pm1\sigma}$} &
  \multicolumn{1}{c}{$M_J^{\mathrm{mean}\pm1\sigma}$} &
  \multicolumn{1}{c}{$M_K^{\mathrm{mean}\pm1\sigma}$} &
  \multicolumn{1}{c}{$\log \, \left( \frac{M_{\star}}{\mathrm{M_{\sun}}} \right)^{\mathrm{mean}\pm1\sigma}$} &
  \multicolumn{1}{c}{\begin{tabular}[c]{@{}c@{}} SFR\textsuperscript{(16\%,50\%,84\%)}\\ $\rm (M_{\sun} {yr}^{-1})$\end{tabular}} &
  \multicolumn{1}{c}{\begin{tabular}[c]{@{}c@{}}sSFR \textsuperscript{(16\%,50\%,84\%)}\\ $\rm (\times \num{e-12} \, {yr}^{-1})$\end{tabular}}
  \\ 
\midrule 
  $\mathcal{A}1$ & $-18.78 \pm 0.84$ & $-20.00 \pm 0.87$ & $-20.58 \pm 0.93$ & $-21.41 \pm 1.03$ & $-21.43 \pm 1.07$ & $10.18 \pm 0.50$ & $(0.04, 0.69, 2.50)$ & $(1.45, 63.50, 336.40)$\\
  $\mathcal{B}1 $ & $-19.46 \pm 0.74$ & $-20.88 \pm 0.69$ & $-21.59 \pm 0.67$ & $-22.57 \pm 0.65$ & $-22.63 \pm 0.65$ & $10.75 \pm 0.27$ & $(0.02, 0.26, 2.91)$ & $(0.42, 5.35, 73.80)$\\
  $ \mathcal{B}2 $ & $-19.84 \pm 0.62$ & $-21.25 \pm 0.58$ & $-21.95 \pm 0.58$ & $-22.91 \pm 0.59$ & $-22.96 \pm 0.60$ & $10.88 \pm 0.28$ & $(0.03, 0.43, 3.90)$ & $(0.42, 5.72, 90.69)$\\
  $\mathcal{C}1$ & $-20.00 \pm 0.65$ & $-21.49 \pm 0.58$ & $-22.24 \pm 0.55$ & $-23.24 \pm 0.53$ & $-23.30 \pm 0.54$ & $11.05 \pm 0.22$ & $(0.03, 0.19, 2.05)$ & $(0.29, 2.19, 29.33)$\\
  $\mathcal{C}2$ & $-19.98 \pm 0.63$ & $-21.47 \pm 0.57$ & $-22.23 \pm 0.54$ & $-23.23 \pm 0.52$ & $-23.29 \pm 0.53$ & $11.05 \pm 0.21$ & $(0.03, 0.19, 2.06)$ & $(0.29, 2.14, 28.02)$\\
  $\mathcal{C}3$ & $-20.02 \pm 0.59$ & $-21.52 \pm 0.53$ & $-22.27 \pm 0.52$ & $-23.26 \pm 0.51$ & $-23.32 \pm 0.52$ & $11.06 \pm 0.21$ & $(0.023, 0.20, 2.32)$ & $(0.29, 2.10, 29.38)$\\
  $\mathcal{C}4$ & $-19.73 \pm 0.59$ & $-21.53 \pm 0.35$ & $-22.29 \pm 0.35$ & ... & $-23.33 \pm 0.39$ & $11.16 \pm 0.23$ &$(0.0, 0.0, 0.0)$ & $(0.0, 0.0, 0.0)$\\
  $\mathcal{C}5$ & $-19.86 \pm 0.52$ & $-21.63 \pm 0.35$ & $-22.39 \pm 0.38$& ... & $-23.47 \pm 0.39$  & $11.19 \pm 0.25$ & $(0.0, 0.0, 0.0)$ & $(0.0, 0.0, 0.0)$\\
\bottomrule
\end{tabular}
}
\end{center}
\tablefoot{The columns represent the sample label, mean absolute magnitudes in the $u, g, r, J,$ and $K$ bands, mean stellar mass, and 16-, 50- (median), and 84-percentiles of SFR and sSFR of the corresponding sample. 
The uncertainty with each property mean represents its standard deviation within the sample.}
\label{table:subsamples_properties}
\end{table*}

\subsection{SDSS samples for comparison}\label{sec:data_sdss}

\begin{figure*}[t]
    \centering
    \includegraphics[width=\linewidth]{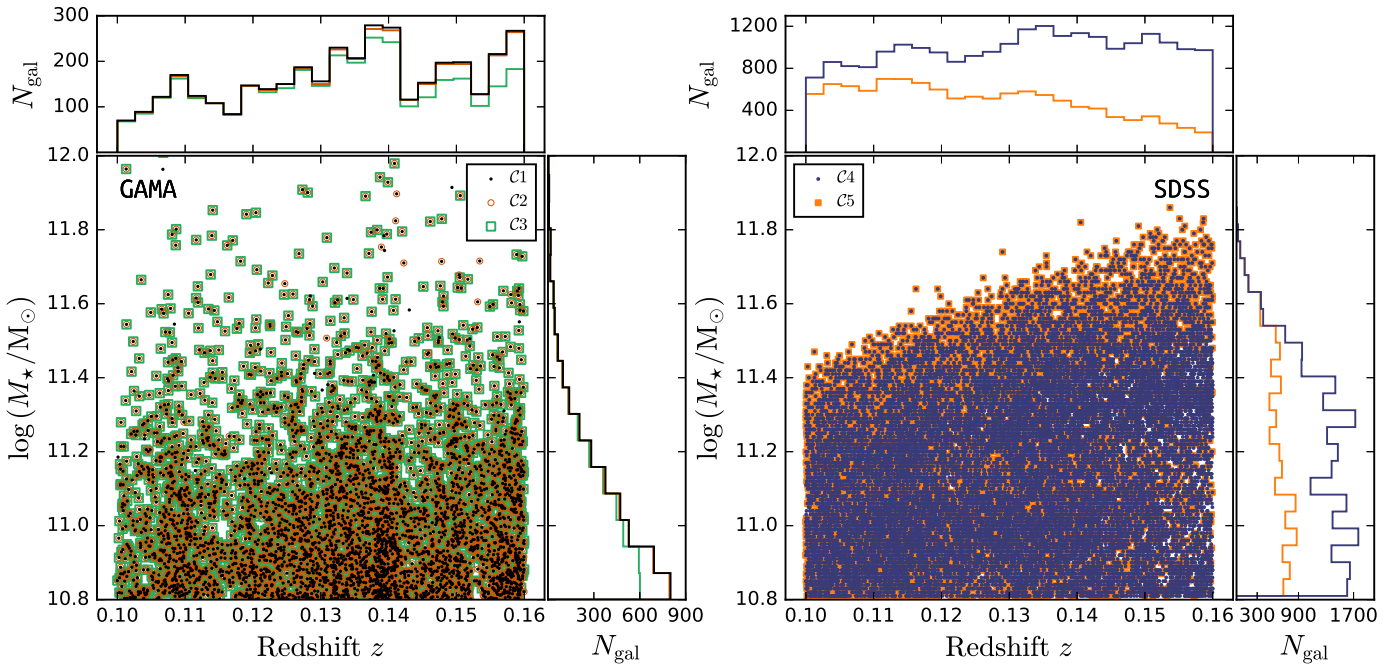}
    \caption{Redshift and stellar mass distribution of stellar-mass-selected samples mentioned in Table~\ref{table:subsamples_definitions}. 
    The left panel represents the $\mathcal{C}1$ ($r < 19.8$; black dots), $\mathcal{C}2$ ($r < 18.8$; red circles), and $\mathcal{C}3$ ($r < 17.8$; green squares) galaxy samples from GAMA and the right panel shows the $\mathcal{C}4$ ($r < 17.8$; blue dots) and $\mathcal{C}5$ ($r < 16.8$; orange squares) galaxy samples from SDSS.
    The top and right histograms of both the panels show the distribution of redshift and stellar mass, respectively.}
    \label{fig:data_c1_c5}
\end{figure*}

Apart from the comparisons between $\mathcal{B}1$ and $\mathcal{B}2$, we also compare the results with the SDSS \citep[][]{york2000_sdss_gen} to understand better how brighter flux limits can affect our measurements.
We use the LSS catalogue and the corresponding random catalogue\footnote{https://data.sdss.org/sas/dr12/boss/lss/} generated from SDSS III Baryon Oscillation Spectroscopic Survey \citep[BOSS;][]{dawson2013_sdssboss} Data Release 12 \citep[DR12;][]{alam2015_sdss_dr12}. 
The SDSS BOSS DR12 catalogue encompasses massive galaxies partitioned into two non-overlapping redshift bins named as ‘LOWZ’ and ‘CMASS’ that cover galaxies in the redshift ranges $z < 0.43$ and $z > 0.43$, respectively.
\citet{reid2016_sdssdr12lss} describe the methods used in the target selection of the SDSS galaxy data sets, and give details of the MKSAMPLE code used to create the LSS catalogue and random catalogue. 
The total sky coverage of the LOWZ DR12 sample is 8337.47 deg\textsuperscript{2}.
In this work, we make use of LOWZ galaxies in the North Galactic Cap in the redshift range $0.1 < z < 0.16$. 

All the galaxies in the LSS catalogue are assigned the $r$-band apparent magnitude from \textsc{SpecPhotoAll} table by matching the angular position within 2\arcsec.
Stellar masses are then assigned by cross-matching with the table \textsc{stellarMassStarformingPort} using \texttt{specObjID}.
The stellar masses are estimated from the best-fit SED obtained from the stellar population model of \citet{maraston2009}.
The fits are performed on the observed $ugriz$-magnitudes of BOSS galaxies with the spectroscopic redshift determined using an adaptation of Hyper-Z code of \citet{bolzonella2000}.
The magnitudes used are extinction-corrected \texttt{model} magnitudes that are scaled to the $i$-band \texttt{cmodel} magnitude.
The GAMA and SDSS stellar masses are derived using different photometry.
Despite the systematic differences between SDSS and GAMA photometry, we find relatively good agreement between GAMA and SDSS stellar masses of overlapping galaxies, taking into account that the common sample is rather small (388 objects). 
The median offset between the masses is $0.18$ dex with a scatter of the order of 0.2 dex.
Hence we use the same stellar mass cuts in GAMA and SDSS to define samples for comparison.

For better statistics, we fix the brightest magnitude limit in our work to be $r < 16.8$.
For the comparison between GAMA and SDSS, we define five stellar-mass-selected samples with the same stellar mass cut of $\log \, (M_{\star}/\mathrm{M_{\sun}}) > 10.8$, but different flux limits.
This gives samples $\mathcal{C}1$, $\mathcal{C}2$, $\mathcal{C}3$ from our parent sample in GAMA with flux limits of $r < 19.8, 18.8, 17.8$, respectively.
From SDSS, we have samples $\mathcal{C}4$, $\mathcal{C}5$ with flux limit of $r < 17.8, 16.8$, respectively.
More details of these samples are given in Table~\ref{table:subsamples_definitions} and Table~\ref{table:subsamples_properties}, respectively, and their mass distribution is shown in Fig.~\ref{fig:data_c1_c5}.
The SFRs, sSFRs, and $K$-band absolute magnitudes of SDSS samples mentioned in Table~\ref{table:subsamples_properties} are derived from the same SED fits from which SDSS stellar masses are derived.
The absolute magnitudes in the $u, g,$ and $r$ bands of SDSS samples are taken from the \textsc{Photoz} table \citep{beck2016}.
The angular distribution of random galaxies for each SDSS sample are taken from the LSS random catalogue and the redshift is randomly assigned from a smoothened $N(z)$ distribution of the corresponding real galaxy sample. 

\section{Measurement methods}\label{sec:measurement}

\subsection{The galaxy two-point correlation function}\label{sec:measurement_2pcf}

The galaxy 2pCF, $\xi(r)$, is a statistical tool used to measure the clustering of galaxies. 
It is defined as the excess probability above random of observing a pair of galaxies at a given spatial separation $r$ in a volume element $dV$ \citep{peebles1980}, that is
\begin{equation}\label{eqn:dp}
    dP = n \, [ 1 + \xi(r) ] \, dV ,
\end{equation}
where $n$ is the number density of galaxies.

It has been observed that CF mostly follows a power law \citep{groth&peebles1977} given by
\begin{equation}\label{eqn:powerlaw}
\xi(r)=\left(\frac{r}{r_0}\right)^{-\gamma} ,
\end{equation}
where $r_0$ and $\gamma$ are the correlation length and slope, respectively.

In practice, owing to the limitations of galaxy surveys, various estimators of $\xi(r)$ have been proposed to minimise the effects related to the limited number of objects and limited survey areas \citep[e.g.][]{davis&peebles1983, hamilton1993}.
The \citet{landy&szalay1993} estimator is the most widely used because of its capability to minimise the above-mentioned problems and is defined by,
\begin{equation} \label{eqn:landy-szalay} 
    \xi(r)= \frac{\langle DD(r) \rangle - 2 \langle DR(r) \rangle + \langle RR(r) \rangle}{\langle RR(r) \rangle} ,
\end{equation}
where $DD(r)$ is the observed number of galaxy-galaxy pairs with the separation in the bin centred at $r$ in the real galaxy sample, $RR(r)$ is the expected number of such pairs from a random galaxy distribution, $DR(r)$ is the number of cross pairs of galaxies between the real and random sample, and $\langle \rangle$ refers to the quantity normalised by the total number of such pairs. 
The random galaxy sample reflects the same sky distribution and redshift distribution of the real galaxy sample.
The number of random galaxies used for the computation is set to be significantly greater (5-10 times) than the number of real galaxies to avoid shot noise on smaller scales. 

To account for the distortions in CF measurements caused by galaxy peculiar velocities, the co-moving redshift space separation between the galaxies is split into two components: parallel ($\pi$) and perpendicular ($r_\mathrm{p}$) to the line of sight. 
The CF thus takes form of a two-dimensional grid $\xi(r_\mathrm{p},\pi)$. 
Integrating $\xi(r_\mathrm{p},\pi)$ over the line-of-sight ($\pi$) direction gives us the projected 2pCF, $\omega_\mathrm{p}(r_\mathrm{p}),$ which can be used to recover the real space CF devoid of redshift space distortions \citep{davis&peebles1983}. 
It is defined as
\begin{equation}\label{eqn:projectedcf}
    \omega_\mathrm{p}(r_\mathrm{p}) = 2 \, \int_0^{\pi_\text{max}} \xi(r_\mathrm{p},\pi) \, \mathrm d\pi .
\end{equation}
The limit of integration $\pi_\text{max}$ has to be reasonable enough to include all the correlated pairs and reduce the noise in the estimator. 
Following Appendix B of \citet{loveday2018_gama_pvd}, we choose the value of $\pi_\text{max}$ to be $40 \, h^{-1} \mathrm{Mpc}$. 

There have been many studies in GAMA using 2pCF in the past. 
The dependence of projected galaxy clustering on various properties was studied by \citet{farrow2015_gama_cf}. 
\citet{loveday2018_gama_pvd} used 2pCF to measure the pairwise velocity distribution in a set of luminosity-selected samples from GAMA.
The small-scale clustering properties of star-forming galaxies were used by \citet{gunawardhana2018} to study the interactions between galaxies.
\citet{christodoulou2012_gama_angular} used CF as a tool to check the robustness of their photometric redshift estimates.
\citet{jarrett2017} analysed the spatial distribution of mid-infrared Wide-field Infrared Survey Explorer (WISE) sources observed in the G12 equatorial region of GAMA.
Large-scale clustering of radio galaxies in the Very Large Array Faint Images of the Radio Sky at Twenty-cm (FIRST) survey over the GAMA survey area was studied by \citet{lindsay2014}.
The clustering measurements in GAMA were also used by \citet{alam2021} to model the redshift space distortions.
The clustering properties of low-redshift ($z < 0.3$) submillimeter galaxies detected at 250 $\mu m$ in the Herschel-ATLAS \citep{eales2010_herschel-atlas} using the redshift information from GAMA was carried out by \citet{vanKampen2012_gama_herschel}. 
\citet{vanUitert_2018} used angular CF of GAMA galaxies as one of the probes to constrain cosmological parameters.
In our work, we examine how galaxy clustering depends on various galaxy properties, such as luminosities in different passbands, stellar mass, and SFR, using MCF.

\subsection{Marked correlation function}\label{sec:measurement_mcf}

The 2pCF characterises the galaxy clustering. 
It can, and successfully has been (as described in Sect.~\ref{sec:introduction}), used to study clustering dependences on various properties of galaxies. 
This is done by defining the galaxy samples based on the property of interest (e.g. luminosity, colour, or stellar mass). 
However, after the selection and further during the analysis, these properties are left unconsidered and each galaxy is weighted equally during CF measurements. 
On the other hand, marked statistics allows us to study the properties of galaxy clustering by taking the physical properties (called marks) of each galaxy in the sample into account. 
These marks can be discrete or continuous values such as luminosity, colour, stellar mass, SFR, and morphology. \citep{sheth2005_galform_models}.

The MCF allows for the efficient study of the spatial distribution of galaxy properties and their correlation with the environment \citep{skibba2013}. 
The two-point MCF is defined as
\begin{equation}\label{eqn:M(r)}
M(r) = \frac{1 + W(r)}{1 + \xi(r)},
\end{equation}
where $\xi(r)$ is the galaxy 2pCF defined by Eq.~(\ref{eqn:landy-szalay}) and $W(r)$ is the weighted CF obtained with the same estimator, but with pair counts computed by weighting each real galaxy in the pair. That is
\begin{equation} \label{eqn:mcf_landy} 
    W(r)= \frac{\langle WW(r) \rangle - 2 \langle WR(r) \rangle + \langle RR(r) \rangle}{\langle RR(r) \rangle} .    
\end{equation}
We adopt multiplicative scheme for pair weighting, that is
\begin{equation}\label{eqn:WW(r)}
    WW(r) = \sum_{ij} w_i \times w_j ,
\end{equation}
where $w_i$ is the weight of the $i^\text{th}$ galaxy given by the ratio of its mark to the mean mark across the sample. 

The projected two-point MCF is defined as
\begin{equation}\label{eqn:projectedMCF}
    M_\mathrm{p}(r_\mathrm{p}) = \frac{1 + W_\mathrm{p}(r_\mathrm{p})/r_\mathrm{p}}{1 + \omega_\mathrm{p}(r_\mathrm{p})/r_\mathrm{p}} .
\end{equation}
Essentially, MCF at a scale $r$ tells us if galaxies in pairs separated by $r$ tend to have larger or smaller values of their mark than the mean mark in the entire sample \citep{sheth&tormen2004}.

\subsection{Rank-ordered marked correlation function}\label{sec:measurement_rank_mcf}

For a given property, a stronger MCF signal at a certain scale indicates greater probability of finding galaxy pairs for which the given property has a larger value for both the galaxies.
Hence the property that is more dependent on environment is that corresponding to a larger MCF.
However, comparing different MCFs obtained using different properties is not straightforward \citep{skibba2013}.
When computing the MCF in a traditional approach, the value of the physical property of a galaxy is considered as its mark and the CF is directly weighted by the ratio of the given mark to the mean mark of the sample. 
Hence, the amplitude of MCF depends on the distribution of the marks and the variations in their formulation (e.g. log or linear). 
This makes it impossible to directly compare different MCFs measured using different properties as marks if these properties have different distribution or formulation. 
\citet{skibba2013} developed a solution to this problem.
Each galaxy is given a rank based on the relative strength of the value of its property in the sample; that is a galaxy with the lowest value is given the lowest rank and another one with a greater value is given a higher rank.
This is called rank-ordering the marks.
The rank of each galaxy is then used as its mark to weight the CF. 
Since all the ranks have a uniform distribution on $[1, N]$, the amplitudes of the MCFs thus obtained using different properties can be compared . 
However, since the weight is given by the rank rather than the property value, any information contained in the shape of the  distribution of the property is lost. 
As we are interested in relative importance of different properties for correlation measurements, all the MCFs shown in this work are rank ordered.

\subsection{Error estimates}\label{sec:measurement_errors}

Since the galaxies are clustered and the pair counts in different bins of $r_\mathrm{p}$ can include the same galaxies, the values of $\omega_\mathrm{p}$ for different bins are correlated. 
Hence the statistical errors associated with clustering measurements are estimated using the covariance matrix obtained from various methods of internal error estimation. 
In our work, we use the jackknife resampling method \citep{norberg2009}, in which we divide the entire sky region into $N_\text{jk}$ subsamples of equal area.
Then $N_\text{jk}$ different jackknife copies of the parent sample are created by omitting one of these subsamples in turn.
Then the CF is measured in each jackknife copy.
We adopt $N_\text{jk} = 12$, which we found to be an optimal number so that the size of each subsample is larger than the maximum scale at which we measure $\omega_\mathrm{p}(r_\mathrm{p})$ ($\sim$ 10 $h^{-1} \, \mathrm{Mpc}$).

The associated covariance matrix is given by
\begin{equation}\label{eqn:jackknife}
C_{ij} = \frac{N_\text{jk} - 1}{N_\text{jk}} \sum\limits_{k=1}^{N_\text{jk}} \left( \omega_\mathrm{p}^k(r_i) - \bar\omega_\mathrm{p}(r_i) \right) \, \, \left( \omega_\mathrm{p}^k(r_j) - \bar\omega_\mathrm{p}(r_j) \right) ,
\end{equation}
where $\omega_\mathrm{p}^k(r_j)$ represents the measurement of $\omega_\mathrm{p}$ at $r_\mathrm{p} = r_j$ in the $k$th jackknife copy and $\bar{\omega_\mathrm{p}}$ is the average from $N_\text{jk}$ copies. 
The square root of the diagonal elements of $C_{ij}$ gives the error bar for the $\omega_\mathrm{p}$ at the corresponding bin.

We estimate $r_0$ and $\gamma$ (the power-law fit parameters of $\xi(r)$) from the projected function $\omega_\mathrm{p}(r_\mathrm{p})$.
Using the parametrisation given in Eq.~(\ref{eqn:powerlaw}), the integral in Eq.~(\ref{eqn:projectedcf}) can be analytically performed to give the power-law fit parameters as 
\begin{equation}\label{eqn:analytic_wp}
\omega_\mathrm{p}(r_\mathrm{p}) = r_\mathrm{p} \, \left( \frac{r_\mathrm{p}}{r_0} \right)^{-\gamma}  \, \frac{\Gamma \left(\frac{1}{2}\right) \, \Gamma \left(\frac{\gamma-1}{2} \right)}{\Gamma \left(\frac{\gamma}{2}\right)} ,
\end{equation}
where $\Gamma(n)$ is Euler's Gamma function \citep{davis&peebles1983}.

The power-law fit parameters are usually estimated by minimising the generalised $\chi^2$ using the inverse of full covariance matrix \citep{fisher1994, pollo2005}.
But, limited number of jackknife resamplings can introduce noise to the
non-diagonal elements of covariance matrix.
This can lead to unreliable power-law fit parameters. 
Hence, we adopt the fitting procedure previously done on GAMA data by \citet{farrow2015_gama_cf}.
For that, we first normalise the covariance matrix to give the correlation matrix ($\tilde{C}_{ij}$) given by
\begin{equation}\label{eqn:corr_mat}
  \tilde{C}_{ij} = \frac{C_{ij}}{\sigma_i \sigma_j} ,
\end{equation}
where $\sigma_i = \sqrt{C_{ii}}$, the error bar associated with the $\omega_\text{p}$ measurement at the $i$th bin.

The correlation matrix $\mathbf{\tilde{C}}$ is then transformed into a matrix $\mathbf{D}$ using singular value decomposition (SVD) given by $\mathbf{\tilde{C}} = \mathbf{U}^\mathrm{T} \, \mathbf{D} \, \mathbf{U}$.
The diagonal matrix $\mathbf{D}$ has $\uplambda^2_{ij} \, \delta_{ij}$ as elements, where $\delta_{ij}$ is the Kronecker delta.
Columns of the matrix $\mathbf{U}$ are the eigen modes of the correlation matrix, where $\uplambda^2_{ij} \delta_{ij}$ are their corresponding eigen values. 
The inverse of the diagonal matrix $\mathbf{D}$ is given by $D^{-1}_{ij} = (1/\uplambda^2_{ij}) \, \delta_{ij}$.
Then we transform the correlation matrix back using $\mathbf{\tilde{C}^{-1}} = \mathbf{U} \, \mathbf{D}^{-1} \, \mathbf{U}^\mathrm{T}$ and while doing this calculation, we set $D^{-1}_{ii} = 0$ for those eigen values $\uplambda_{ii}^2 < \sqrt{2/N_\text{jk}}$ \citep{gaztanaga2005}. 
This effectively removes the influence of least significant eigen modes that are more likely to suffer from noise. 

The transformed inverse of the correlation matrix is then used to minimise the $\chi^2$, defined as
\begin{equation}\label{eqn:chi_square}
 \chi^2 = \sum \limits_{i,j} \frac{\left( \omega_\mathrm{p}^\text{mod}(r_i) - \omega_\mathrm{p}(r_i) \right)}{\sigma_i} \,\, \tilde{C}^{-1}_{ij} \,\, \frac{\left( \omega_\mathrm{p}^\text{mod}(r_j) - \omega_\mathrm{p}(r_j) \right)}{\sigma_j} ,
\end{equation}
where $\omega_\mathrm{p}(r_i)$ is the measured value of CF at $r_\mathrm{p} = r_i$ and $\omega_\mathrm{p}^\text{mod}$ is the power-law model value given by Eq.~(\ref{eqn:analytic_wp}).
The uncertainties in the power-law parameters $r_0$ and $\gamma$ presented in this work are defined by the 68.3\% joint confidence levels \citep[Chapter 15.6 of][]{press1992}.
\citet{gaztanaga2005} and \citet{marin2013} provide detailed descriptions of the fitting procedure using SVD to estimate the power-law parameters.

As the errors in $W_\mathrm{p}(r_\mathrm{p})$ and $\omega_\mathrm{p}(r_\mathrm{p})$ are strongly correlated, simply summing these  in quadrature gives an overestimate for the error in $M_\mathrm{p}(r_\mathrm{p})$.
A much better approximation of the uncertainty is obtained by randomly scrambling the marks among the galaxies and remeasuring $M_\mathrm{p}$. 
This is repeated $\sim 100$ times and the standard deviation around the mean gives the uncertainty in $M_\mathrm{p}$ \citep{skibba2006}.

\section{Results}\label{sec:results}

\begin{figure*}[t]
    \centering
    \includegraphics[width=\linewidth]{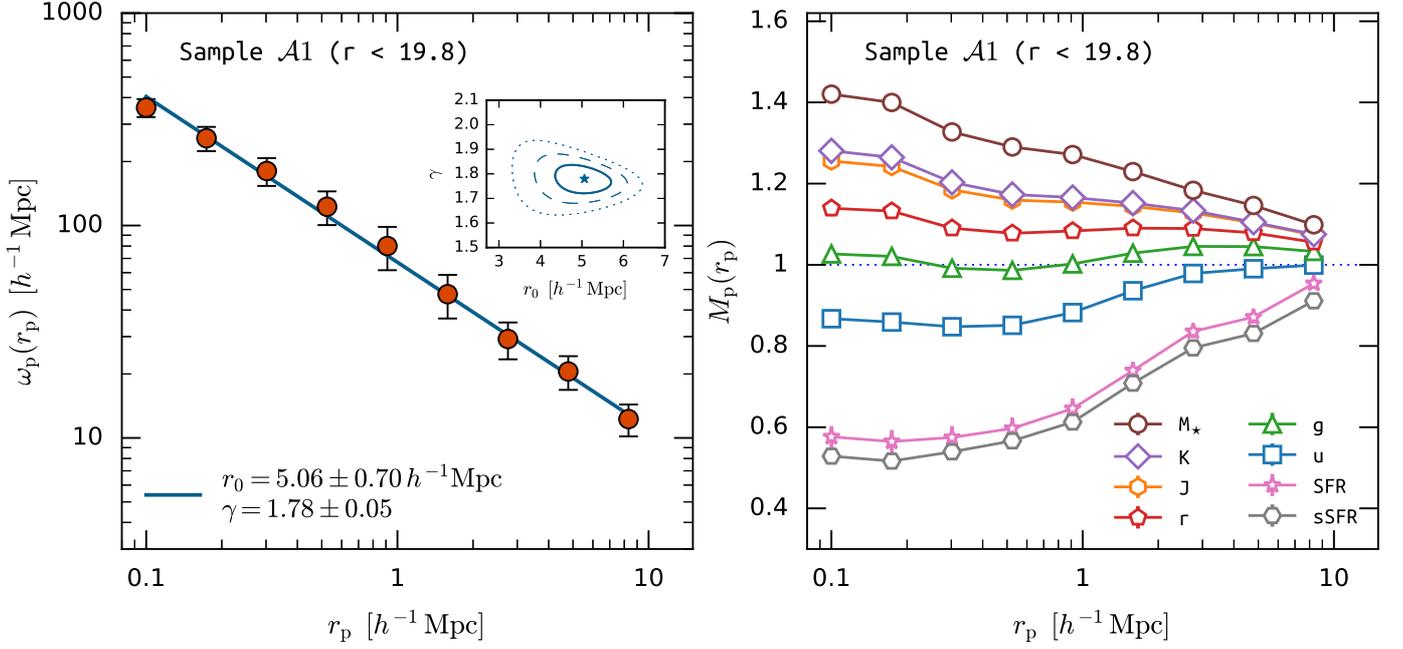}
    \caption{Projected 2pCF $\omega_\mathrm{p}(r_\mathrm{p})$ with a power-law fit (filled markers; left panel) and rank-ordered projected MCFs $M_\mathrm{p}(r_\mathrm{p})$ (unfilled markers; right panel) for Sample $\mathcal{A}1$ described in Sect.~\ref{sec:data_sampleselection}. In the left panel, error bars of $\omega_\mathrm{p}(r_\mathrm{p})$ are the square root of the diagonals of the covariance matrix obtained from the jackknife resampling method. The inset in the left panel shows the power-law fit parameters (filled star) and their 1$\sigma$, 2$\sigma$, and 3$\sigma$ error contours (solid, dashed, and dotted, respectively).
    In the right panel the different symbols represent measurements with different marks (as labelled), and the error bars are obtained by random scrambling of the marks.
    The error bars of $M_\mathrm{p}(r_\mathrm{p})$ are too small to be visible.}
    \label{fig:result_a1}
\end{figure*}

\begin{figure*}[t]
    \centering
    \includegraphics[width=0.9\linewidth]{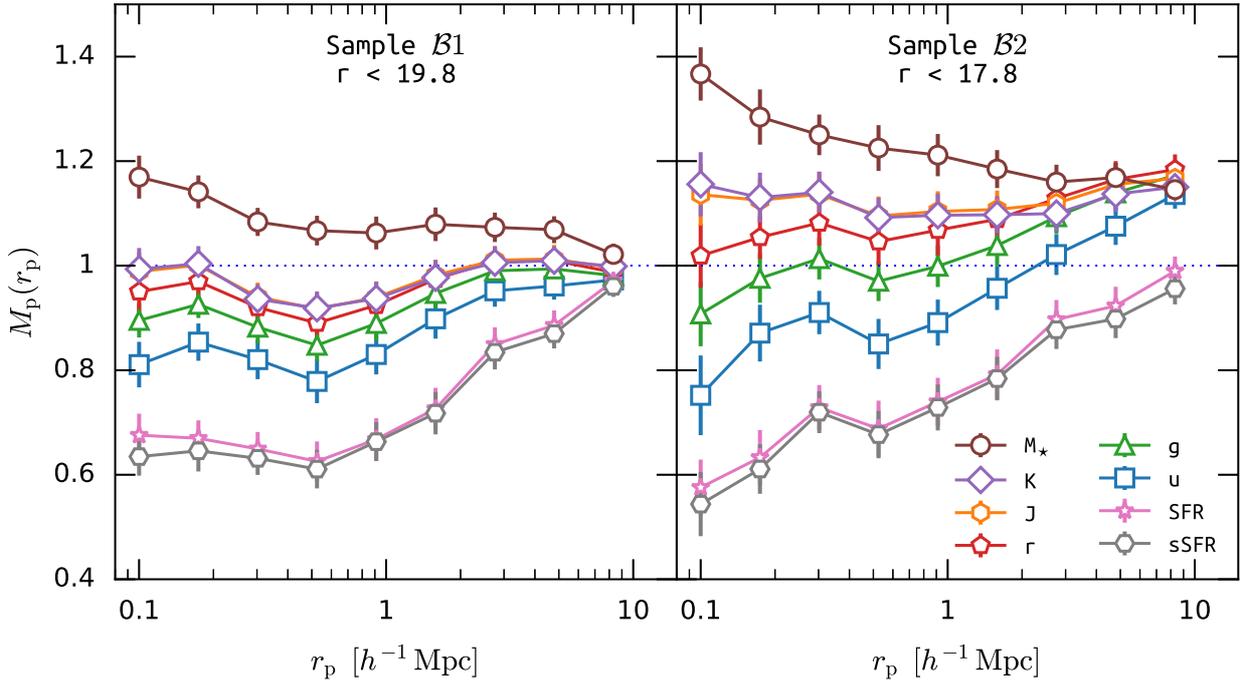}
    \caption{Rank-ordered projected MCFs $M_\mathrm{p}(r_\mathrm{p})$ for samples $\mathcal{B}1$ (left) and $\mathcal{B}2$ (right) with different flux limits, as labelled. The different symbols represent different marks considered for the $M_\mathrm{p}(r_\mathrm{p})$ measurement (as labelled) and the error bars are obtained by random scrambling of the marks.}
    \label{fig:result_b1_b2}
\end{figure*}

\begin{figure*}[h]
    \centering
    \includegraphics[width=\linewidth]{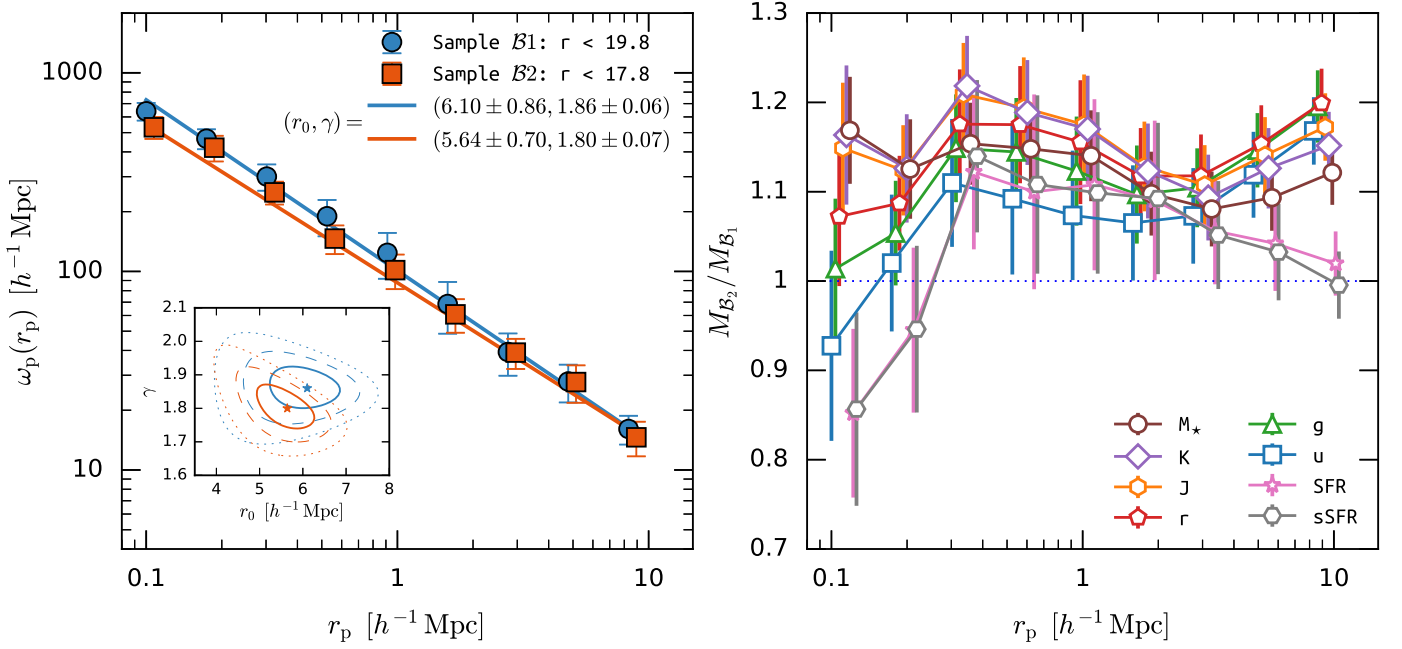}
    \caption{Projected 2pCFs $\omega_\mathrm{p}(r_\mathrm{p})$ for samples $\mathcal{B}1$ and  $\mathcal{B}2$ (filled markers; left panel) and the ratio $M_{\mathcal{B}2}/M_{\mathcal{B}1}$ between MCFs obtained for these two samples (unfilled markers; right panel). The different  symbols in the right panel represent the ratio between  different MCFs measured with corresponding galaxy property chosen as a mark. Small offsets along x-axis have been added for clarity. The error bars for $\omega_\mathrm{p}(r_\mathrm{p})$ are obtained from jackknife resampling method. 
    The inset in the left panel shows the power-law fit parameters (filled stars) and their 1$\sigma$, 2$\sigma$, and 3$\sigma$ error contours (solid, dashed, and dotted, respectively).
    In the right panel, the errors are calculated in quadrature.}
    \label{fig:result_b1_b2_ratio}
\end{figure*}

\begin{figure*}[t]
    \centering
    \includegraphics[width=\linewidth]{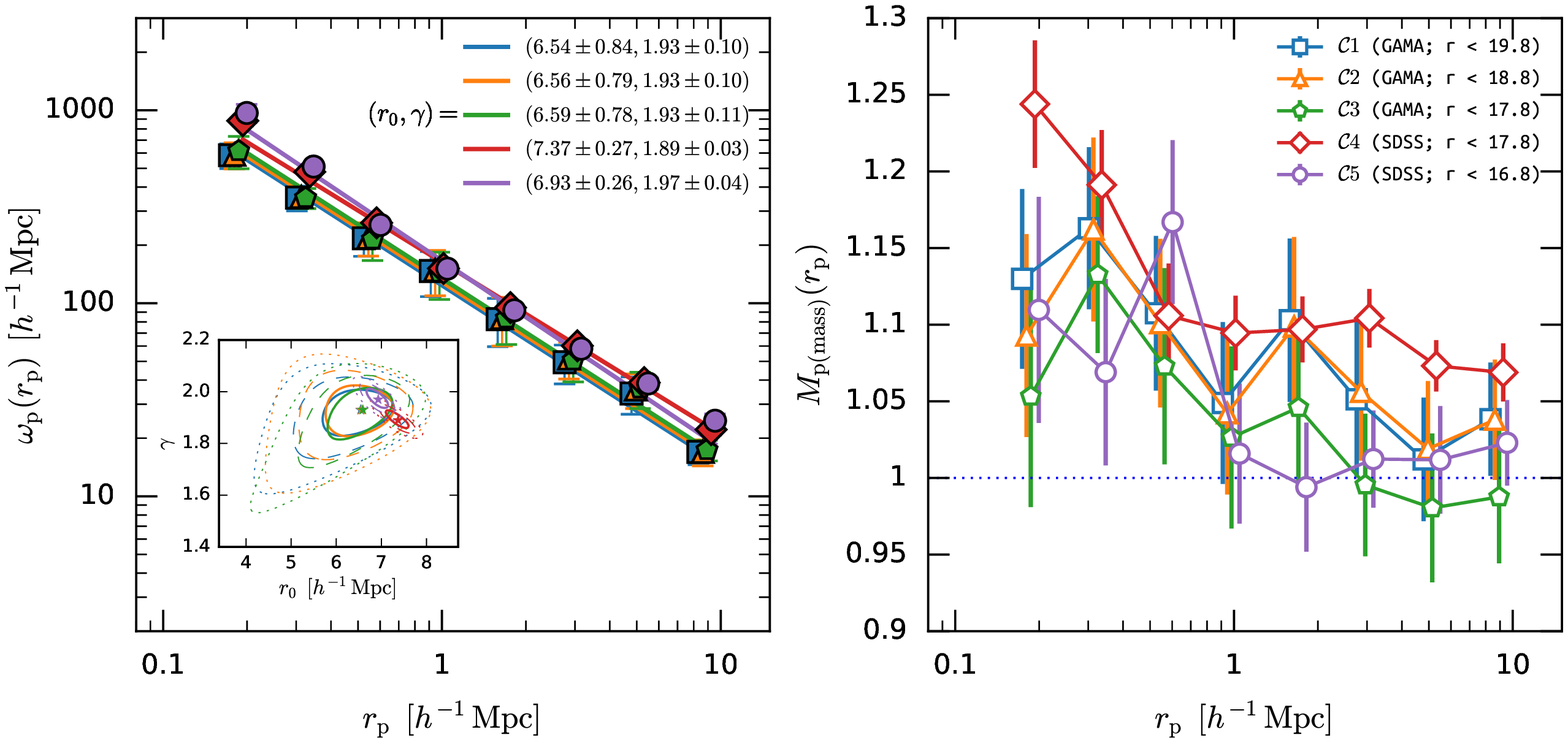}
    \caption{Projected 2pCFs (filled markers; left panel) and stellar mass MCFs (unfilled markers; right panel) in GAMA and SDSS surveys. 
    The different symbols represent the measurements in different samples as labelled.
    The error bars for $\omega_\mathrm{p}(r_\mathrm{p})$ are obtained from jackknife resampling method and those for $M_\mathrm{p}(r_\mathrm{p})$ are obtained by random scrambling of the marks. 
    The inset in the left panel shows the power-law fit parameters (filled stars) and their 1$\sigma$, 2$\sigma$, and 3$\sigma$ error contours (solid, dashed, and dotted, respectively).
    Small offsets along x-axis have been added for clarity.}
    \label{fig:result_c1_c5}
\end{figure*}

In this section, we present our results of the environmental dependence of galaxy luminosity, stellar mass, SFR, and sSFR.
We measure 2pCFs and rank-ordered MCFs for various stellar-mass-selected samples described in Table~\ref{table:subsamples_definitions} selected from GAMA in the redshift range $0.1 < z < 0.16$.
Each 2pCF is fitted with a power-law model and the MCFs are measured using eight different properties as marks: absolute magnitudes in $u, g, r, J,$ and $K$ bands, stellar mass, SFR, and sSFR. 
Details are described in Sect.~\ref{sec:measurement}, especially Eq.~(\ref{eqn:projectedcf}) and (\ref{eqn:projectedMCF}).
In all the samples, we could reliably measure the CFs in the range $0.1 < r_\mathrm{p} < 10 \, h^{-1} \mathrm{Mpc}$. 
The errors in $\omega_\mathrm{p}(r_\mathrm{p})$ are obtained from 12 jackknife realisations and the errors in $M_\mathrm{p}(r_\mathrm{p})$ are obtained by randomising the marks (100 times) as described in Sect.~\ref{sec:measurement_errors}.
The best-fitting power-law parameters for 2pCFs in all the samples are given in Table~\ref{table:wp_params}.

\begin{table}[h]
\caption{Best-fitting power-law parameters for all the galaxy samples used in this work.}
\begin{center}
{\tiny
\begin{tabular}{c c c c c}
\toprule
\toprule
  \multicolumn{1}{c}{$\log \, (M_{\star}/\mathrm{M_{\sun}})^\mathrm{min}$} &
  \multicolumn{1}{c}{Sample} &
  \multicolumn{1}{c}{Flux limit} &
  \multicolumn{1}{c}{$r_0 \, (h^{-1} \mathrm{Mpc})$} &
  \multicolumn{1}{c}{$\gamma$} \\
\midrule
  9.3 & $\mathcal{A}1 $ & $r < 19.8$ & $5.06 \pm 0.70$ & $1.78 \pm 0.05$ \\
  \midrule
  \multirow{2}{*}{10.4} & $\mathcal{B}1 $ & $r < 19.8$ & $6.10 \pm 0.86$ & $1.86 \pm 0.06$ \\
   & $ \mathcal{B}2 $ & $r < 17.8$ & $5.64 \pm 0.70$ & $1.80 \pm 0.07$ \\
  \midrule
  \multirow{5}{*}{10.8} & $\mathcal{C}1$ & $r < 19.8 $ & $6.54 \pm 0.84$ & $1.93 \pm 0.10$ \\
  & $\mathcal{C}2$ & $r < 18.8 $ & $6.56 \pm 0.79$ & $1.93 \pm 0.10$ \\
  & $\mathcal{C}3$ & $r < 17.8 $ & $6.59 \pm 0.78$ & $1.93 \pm 0.11$ \\
  & $\mathcal{C}4$ & $r < 17.8 $ & $7.37 \pm 0.27$ & $1.89 \pm 0.03$   \\
  & $\mathcal{C}5$ & $r < 16.8 $ & $6.93 \pm 0.26$ & $1.97 \pm 0.04$   \\
\bottomrule
\end{tabular}
}
\end{center}
\label{table:wp_params}
\end{table}

\subsection{Two-point and marked correlation functions}\label{sec:results_lum_mass_mcf}

In Fig.~\ref{fig:result_a1} we show the 2pCF and MCFs obtained for galaxies with a flux limit of $r < 19.8$ (Sample $\mathcal{A}1$). 
The left panel shows the projected 2pCF $\omega_\mathrm{p} (r_\mathrm{p})$, which at first approximation obeys a power-law model.
The best-fit parameters are correlation length $r_0 = 5.06 \pm 0.70 \, h^{-1} \mathrm{Mpc}$ and slope $\gamma = 1.78 \pm 0.05$.
This can be compared with the results of \citet{farrow2015_gama_cf}, although their samples vary from ours. 
To be compared with our sample $\mathcal{A}1$, the most appropriate sample of theirs is the sample with mass limits $10 < \log M_{\star}/\mathrm{M_{\sun}} \, h^{-2} < 10.5$ in the redshift range $0.14 < z < 0.18$.
The parameter values for that sample were measured to be $r_0 = 5.94 \pm 0.46 \, h^{-1} \mathrm{Mpc}$ and $\gamma = 1.86 \pm 0.05$.
The correlation length varies by a factor of $1.1\sigma$ from our result.
This small deviation in correlation length is expected as their stellar mass cuts vary from ours.  

The right panel of Fig.~\ref{fig:result_a1} presents rank-ordered MCFs $M_\mathrm{p}(r_\mathrm{p})$ obtained for different galaxy properties used as marks (as described in the legend). 
The presented MCFs strongly deviate from unity on small scales ($r_\mathrm{p} < 1 \, h^{-1} \mathrm{Mpc}$) for all luminosity, stellar mass, SFR, and sSFR marks.
This deviation then decreases, but still remains at larger scales.
In general, a stronger deviation of MCF from unity, means stronger correlation of the corresponding galaxy property with the environment (as described in Sect.~\ref{sec:measurement_rank_mcf}).
As shown in Fig.~\ref{fig:result_a1}, stellar mass MCF deviation from unity is greater than any of the luminosity MCFs.
This means that stellar mass catches the galaxy over-density in small scales better than other properties used in this work.
Thus, stellar mass can be considered as a more direct tracer of environment than luminosity and SFR.

Among the MCFs measured using luminosities in different passbands, the $K$-band MCF has the highest amplitude and the $u$-band MCF has the lowest. 
This means that the $K$-band luminosity traces the environment better than any other band of those used in this work.
Moreover, the $K$-band MCF shows similar, although not exactly the same, behaviour as the stellar mass MCF; the $K$-band luminosity and stellar mass are therefore correlated with environment in a similar fashion. 
This, in turn, confirms that $K$-band luminosity can be used as a tracer (or a proxy) of stellar mass when we lack the measurements of stellar mass \citep{kochanek2001, baldry2012}.  

On the other hand, SFR and sSFR MCFs behave opposite to the stellar mass MCF;  this presents low values (high deviation from unity) on small scales and high values (close to unity) on large scales.
This means that there is only a small number of close pairs of galaxies with strong star formation activity. 
That is the densest regions of the local universe are mainly populated by old or quiescent galaxies.
Similar, although weaker, behaviour is presented by the $u$-band MCF. 
This suggests that the $u$-band luminosity traces galaxy SFRs \citep[or at least serves as a proxy;][]{madau2014}.

\subsection{Dependence on flux limit}\label{sec:results_flux_dep}

To study the impact of survey flux limit on the correlations observed in Sect.~\ref{sec:results_lum_mass_mcf}, we select two distinct samples with the same stellar mass cut $\log \, (M_{\star}/\mathrm{M_{\sun}}) > 10.4$, but different flux limits: sample $\mathcal{B}1$ with $r < 19.8$ and sample $\mathcal{B}2$ with $r < 17.8$ (see Table~\ref{table:subsamples_definitions} for details).
In Fig.~\ref{fig:result_b1_b2}, we show the measurements of MCFs for both of these galaxy samples.
A direct comparison of results is shown in Fig.~\ref{fig:result_b1_b2_ratio}, where we present a 2pCFs (left panel) and the ratio between the MCFs (right panel) measured for sample $\mathcal{B}2$ and sample $\mathcal{B}1$ using different galaxy properties as marks.

For the same flux limit ($r < 19.8$), sample $\mathcal{B}1$ exhibits a larger correlation length than $\mathcal{A}1$, although within $1\sigma$ (see Table~\ref{table:wp_params}). 
By definition, $\mathcal{A}1$ contains many more less massive galaxies than $\mathcal{B}1$.
Our observation that $\mathcal{B}1$ shows slightly stronger clustering than $\mathcal{A}1$ therefore agrees with the common observation that more massive galaxies exhibit stronger clustering than less massive ones \citep[e.g.][]{skibba2015, durkalec2018}.
We also observe in the left panel of Fig.~\ref{fig:result_b1_b2} that the stellar mass and $g, r, J,$ and $K$-band MCFs of $\mathcal{B}1$ sample exhibit lower amplitude than those in $\mathcal{A}1$, with $g, r, J,$ and $K$-band MCFs falling below unity.

When it comes to sample $\mathcal{B}2$, most of the MCFs show a higher amplitude relative to $\mathcal{B}1$ (as shown in the Fig.~\ref{fig:result_b1_b2}).
This can also be observed in the right panel of Fig.~\ref{fig:result_b1_b2_ratio} in which the ratio of MCFs of $\mathcal{B}2$ to that of $\mathcal{B}1$ are above unity on most of the scales.
This can be associated with the stellar mass incompleteness effect and Sect.~\ref{sec:discussion_massincompleteness} deals with a discussion on this effect. 
The stellar mass can still be used as a good indicator of galaxy environment; MCF, with the stellar mass used as mark, exhibits stronger deviation from unity than MCF with any of the luminosity marks.

However, the change in the flux limit does not affect our 2pCF measurements. 
The correlation lengths of samples $\mathcal{B}1$ and $\mathcal{B}2$ agree within $1\sigma$. 
The best-fit power-law parameters for the sample $\mathcal{B}1$ take the values of $r_0 = 6.10 \pm 0.86 \, h^{-1} \mathrm{Mpc}$ and $\gamma = 1.86 \pm 0.06$, whereas the same  parameters for sample $\mathcal{B}2$ are $r_0 = 5.64 \pm 0.70 \, h^{-1} \mathrm{Mpc}$ and $\gamma = 1.80 \pm 0.07$.

\subsection{Comparison with the SDSS}

To further check the effect of survey flux limit on clustering measurements, we extend our studies to even lower magnitude cuts. 
For that we use data from the SDSS survey.
We select a total number of five samples: three from GAMA ($\mathcal{C}1, \mathcal{C}2, \mathcal{C}3$) and two from SDSS ($\mathcal{C}4, \mathcal{C}5$). 
Each of these samples have the same stellar mass cut ($\log \, (M_{\star}/\mathrm{M_{\sun}}) > 10.8$), but different flux limits.
The details are given in Table~\ref{table:subsamples_definitions}. 
Figure~\ref{fig:result_c1_c5} shows the results of 2pCF and MCF (with stellar mass used as mark) measurements for each of these samples.
Additionally, the best-fit power-law parameters for each 2pCF are given in Table~\ref{table:wp_params}.

The correlation lengths obtained for the GAMA samples $\mathcal{C}1$, $\mathcal{C}2$, and $\mathcal{C}3$ are comparable within $1\sigma$.
In case of MCFs, the GAMA samples ($\mathcal{C}1$, $\mathcal{C}2$ and $\mathcal{C}3$) agree between each other within the error bars.
However, the stellar mass MCFs of samples $\mathcal{C}3$ and $\mathcal{C}4$ (with the same flux limit and stellar mass limit) are noticeably different from each other on most of the scales although their correlations lengths agree with each other.
We also observe significant differences between correlation lengths and MCFs of the SDSS samples $\mathcal{C}4$ and $\mathcal{C}5$.

\section{Discussion}\label{sec:discussion}

\subsection{Environmental dependence of galaxy properties}\label{sec:discussion_general}

The MCF is a useful tool to study the environmental dependence of galaxy properties. 
Empirically, it is a ratio of terms involving the weighted and the unweighted CF. 
For a given galaxy property, all galaxy pairs in a sample carry a weight, that is the product of property values for both galaxies in units of its mean value. 
At each spatial scale, the MCF signal depends on the value of the weights at that scale; the larger amplitude of rank-ordered MCF, measured using a particular galaxy property, implies it has a stronger correlation with environment.

As described in Sect.~\ref{sec:results_lum_mass_mcf} and shown in Fig.~\ref{fig:result_a1}, we observe different amplitudes for MCFs measured using different galaxy properties.
All galaxy properties (luminosities, stellar mass, SFR, and sSFR) correlate with the environment, each in different way.
For example, the stellar mass and luminosity (in the $g, r, J,$ and $K$ bands) MCFs take values higher than unity on small scales, indicating an abundance of close galaxy pairs with these property values greater than the average of the sample.
This agrees with the well-known observation that the most  massive and luminous galaxies (in $g, r, J,$ and $K$ bands) are mostly found in dense regions \citep{norberg2002, coil2006, pollo2006, meneux2009, bolzonella2010, abbas2006, abbas2010, zehavi2011, marulli2013, skibba2014_combining_fields, farrow2015_gama_cf, durkalec2018, cochrane2018}. 
This phenomenon can be explained in the framework of the hierarchical structure formation \citep{press_schechter_1974, white&rees1978,mo1996,springer2005}.

\subsection{$u$-band luminosity and star formation rate dependence on the environment}\label{sec:discussion_uband_sfr}

We observe that the $u$-band marked correlation shows different behaviour in comparison to other passbands ($g, r, J,$ and $K$, see Fig.~\ref{fig:result_a1}).
The $u$-band MCF takes values smaller than unity on scales $r_\mathrm{p} < 3 \, h^{-1} \mathrm{Mpc}$, indicating low probability of finding pairs of two galaxies similarly bright in this band.
This is in complete opposition to the results from other bands, where the most $g,r,$ and $K$ luminous galaxies were the most strongly clustered.
This special behaviour of the $u$-band MCF has also been observed by \citet{deng2012} and it agrees with the semi-analytical galaxy formation models \citep[see Fig.~2 of][]{sheth2005_galform_models}.

The $u$-band and UV light are thought to be primarily emitted by starburst galaxies with young stellar populations \citep{cram1998}. 
The UV-selected galaxies exhibit low clustering in the local universe \citep{heinis2004, heinis2007, milliard2007}.
\citet{heinis2004} measured a correlation length of $r_0= 3.2^{+0.8}_{-2.3} \, h^{-1} \mathrm{Mpc}$ for low-redshift UV galaxies from the FOCA survey and \citet{milliard2007} found it to be $ 3.7 \pm 0.6 \, \text{Mpc}$ from the GALEX survey.
Both these observations confirm the weak clustering of UV-selected galaxies in the local universe.
This result agrees with the work done by \citet{barsanti2018} using GAMA groups, in which they found a rise in star formation in galaxies that are located away from the group centre compared to those in the central regions.

Similarly, SFR and sSFR MCFs also have values smaller than unity on all scales. 
The similar behaviour of SFR and sSFR MCFs show that both these properties correlate with environment in a similar way.
This means that active star-forming galaxies are rarely found in close proximity to each other.
This observation can be connected to the known fact that, at low redshifts, active star formation takes place mainly in low-density environments \citep{lewis2002, gomez2003}: less evolved (young) galaxies that formed in less dense regions exhibit strong star formation activity.
Our observations agree with the SFR MCF measurements by \citet{sheth2005_galform_models} and, to some extent, with recent GAMA studies by \citet{gunawardhana2018}; their results show however weaker deviation of SFR and sSFR MCF from unity than ours.
This can be due to the apparent absence of the SFR-density relationship as a consequence of selecting the star-forming sample of galaxies \citep{mcgee2011, wijesinghe2012}.
\citet{gunawardhana2018} measurements were made for samples of actively star-forming galaxies.

The similar behaviour of $u$ band and SFR MCFs has a practical interpretation.
The SFR of a galaxy can be estimated by applying a scaling factor to the luminosity measurements sensitive to star formation \citep{condon1992, kennicutt1998, madau2014}. 
Since the $u$-band light is dominated by starburst galaxies with young stellar populations, it is more closely correlated to SFR than to stellar mass in galaxies \citep{hopkins2003} and is hence considered to be an indicator of SFR.
This correlation is reflected in our results; $u$-band MCF follows SFR MCF in tracing the galaxy environment at small scales.
This suggests that $u$-band luminosity can be a good proxy of SFR in the context of galaxy clustering.

\subsection{The most reliable tracer of galaxy environment}\label{sec:discussion_bettertracer}

The amplitude of rank-ordered MCFs, computed using various marks, can be used to find the galaxy property with the strongest environmental dependence.
Our observations suggest that this parameter is the stellar mass.
We therefore agree with the past results that the distribution of massive galaxies is strongly correlated with dark matter over-densities \citep{kauffmann2004, scodeggio2009, davidzon2016}. 
This dependence is expected from the hierarchical structure formation theory according to which the local density contrast is connected to the properties of the hosting halo, which is further related to the galaxy stellar mass \citep{moster2010, wechsler2018}.
The strong correlation between stellar mass and environment has been taken into account in studies that explore the environmental dependence of galaxy properties \citep[e.g.][]{kauffmann2004, peng2010}.

Our observations can be also interpreted in terms of galaxy evolution.
Stellar mass plays an important role in shaping galaxy star formation history \citep{gavazzi1996, kauffmann2003, heavens2004}, which is yet another parameter that is strongly correlated with the  local environment of a galaxy \citep{kauffmann2004, blanton2005}.

It is important to note a feature of our studies that might influence our results, namely the galaxy sample selection.
In this work we measure MCFs in stellar mass selected samples only.
That is our samples are nearly complete only in stellar mass and not in other properties.
Our observation -- stellar mass MCF being more enhanced than other MCFs -- could be an outcome of this selection.
However, in our preliminary analysis \citep{sureshkumar2020}, we measured the same MCFs, for $u, g, r,$ and $K$-band luminosities, and stellar mass marks, using samples selected based on the corresponding property.
We observed similar trends as here, ruling out the effect of sample selection on our observation.

\subsection{$K$-band luminosity as a proxy for stellar mass}\label{sec:discussion_massproxy}

In our work we also studied the behaviour of MCFs measured using different photometric wavebands and we observe clear differences between MCFs marked with $u, g, r, J,$ and $K$ luminosities (see Fig.~\ref{fig:result_a1}).
This means that luminosity in different passbands correlates differently with environment.
Most significantly, the $K$ band (the reddest considered) shows a stronger environmental dependence than bluer bands. 
This means that there is higher probability to find close pairs of galaxies similarly luminous in the $K$ band than in other bands.
In other words, galaxies luminous in the $K$ band are strongly clustered. 
This observation is in agreement with various clustering studies \citep[e.g.][]{oliver2004, torre2007}.
In particular, \citet{sobral2010}, using sample of H$\alpha$ emitters from HiZELS survey, show a strong increase in galaxy clustering with increasing $K$-band luminosity, but a weak trend in case of the $B$ band.
This difference in clustering strength between various bands is reflected in the varying amplitude of MCFs in Fig.~\ref{fig:result_a1}.

Comparing the amplitudes of stellar mass and $K$-band MCFs, we observe that these galaxy properties trace the environment in a similar way.
This means that the $K$-band luminosity can be used as the second-most reliable galaxy property (among those considered here), after galaxy stellar mass, to trace the environment.
In other words, a sample that is complete in $K$-band luminosity can be a good substitute of a stellar-mass-complete sample.

This observation agrees with the existing results.
It has been shown that longer-wavelength luminosities (e.g. $K$ band) are dominated by evolved stellar populations and are least affected by dust extinction, making them directly related to the galaxy stellar mass \citep{cowie1994, gavazzi1996, kauffmann1998, kochanek2001, baldry2012, jarrett2013, cluver2014}.
\citet{kauffmann1998} pointed out that infrared light is a much more robust tracer of stellar mass than optical light out to $z \sim 1 - 2$.
They observed that galaxies of the same stellar masses have the same $K$-band luminosities, independent of their star formation histories.
Additionally, near-IR (mainly $K$-band) luminosity functions were well utilised to estimate stellar mass distribution in the local universe \citep{cole2001, kochanek2001, bell2003, drory2004, zhu2010, meidt2014}.
Using a matched GAMA-\textit{WISE} catalogue, \citet{cluver2014} explored the usability of mid-IR wavelengths (W1 and W2) for stellar mass estimation.

However, the relation between $K$-band luminosity and stellar mass is not entirely direct \citep{vanderwel2006, kannappan2007}.
In our results, we also observe the differences between stellar mass and $K$-band MCFs on small scales (see Fig.~\ref{fig:result_a1}).
The close galaxy pairs ($r_\mathrm{p}<1 \, h^{-1} \mathrm{Mpc}$) show stronger signals when weighted using stellar mass than when weighted with $K$-band luminosity.
Given the fact that the latter is proportional to stellar mass, one possible reason for the difference between both MCFs could be the environmental dependence of the correlation between stellar mass and $K$-band luminosity.
This is in qualitative agreement with the predictions of semi-analytic galaxy formation models proposed by \citet{sheth2005_galform_models}.
Their measurements were made on a volume-limited sample with $z = 0.2$ with a limiting stellar mass of $2 \times 10^{10} \, h^{-1} \mathrm{M_{\sun}}$ that corresponds to $\log \, (M_{\star}/\mathrm{M_{\sun}}) > 10.45$ in our cosmological model. 
Using the observed difference between $K$-band and stellar mass MCFs (see Fig.~5 of \citet{sheth2005_galform_models}), they concluded that the correlation between both these properties depends on environment.
The difference in $K$-band and stellar mass MCFs could also be the result of the stronger correlation between stellar mass and halo mass with respect to $K$-band luminosity \citep{moster2010}.
Hence the stellar mass MCF picks up the environmental dependence of halo mass better than $K$-band MCF.

There is yet another caveat in using $K$-band luminosity selected samples as a proxy for stellar mass selection.
By using this approximation the most evolved, red galaxies in their sample are missed, especially on small scales.
\citet{cochrane2018} observed that $K$-band derived stellar masses are underestimated with respect to full SED stellar masses.
This means that some red galaxies might wrongly end up below the applied stellar mass cut and not be selected when $K$-band stellar mass approximation is used.
This is extremely important for high-detail clustering studies. 
It has been shown that red galaxies tend to occupy dense environments \citep[see e.g.][]{zehavi2005, coil2008, zehavi2011, palamara2013}.
Samples unrepresented owing to the $K$-band selection might, therefore, show weaker than actual clustering properties.   

\subsection{Mass incompleteness effect of flux-limited galaxies}\label{sec:discussion_massincompleteness}

Samples $\mathcal{B}1$ and $\mathcal{B}2$ have the same stellar mass limit, but different flux limits in the $r$ band.
Namely, sample $\mathcal{B}1$ goes fainter with $r<19.8$ and sample $\mathcal{B}2$ is brighter with $r<17.8$ (see Table~\ref{table:subsamples_definitions} for details).
As a result of the imposed flux limit, sample $\mathcal{B}2$ is partially incomplete and significantly less numerous than the complete sample $\mathcal{B}1$; sample $\mathcal{B}2$  lacks galaxies that have mass $\log \, (M_{\star}/\mathrm{M_{\sun}}) > 10.4$ but are not luminous enough to cross the flux limit of $r < 17.8$ (Fig.~\ref{fig:data_b1_b2}).

From the measurements of 2pCFs, we observe that the correlation length does not differ significantly between $\mathcal{B}1$ and $\mathcal{B}2$ (see left panel of Fig.~\ref{fig:result_b1_b2_ratio}).
On the other hand, the MCFs are affected by this difference.
In the right panel of Fig.~\ref{fig:result_b1_b2_ratio}, we show the ratio of MCFs between samples $\mathcal{B}2$ and $\mathcal{B}1$.
It is to be noted that correlation studies based on the $u$-band luminosity are influenced at small scales by the imposed $r$-band flux thresholds.
A difference between these two samples is visible in the $u$-band MCF on the smallest scale ($r_\mathrm{p}\sim0.1 \, h^{-1} \mathrm{Mpc}$) in the right panel of Fig.~\ref{fig:result_b1_b2_ratio}, where a brighter sample shows a weaker MCF signal. 
This means that close pairs of galaxies that are similarly brighter in the $u$ band drop out from the sample with the lower magnitude limit.
The same behaviour is reflected in the SFR and sSFR MCFs.
This is an important observation because the $u$-band luminosity is a very good tracer of star formation processes.
So, even though samples $\mathcal{B}1$ and $\mathcal{B}2$ have the same stellar mass limit, clustering studies on a brighter flux-limited sample ($r<17.8$) lose information about starburst galaxies.
This effect has to be properly corrected for, for example by using methods discussed in \citet{meneux2008, meneux2009}.

As mentioned in Sect.~\ref{sec:results_flux_dep}, we observe a lowering of most of the MCF amplitudes while shifting from sample $\mathcal{A}1$ to $\mathcal{B}1$. 
Interestingly, we also observe a rise in the amplitudes while shifting from $\mathcal{B}1$ to $\mathcal{B}2$, particularly at the larger scales. 
It is to be noted that, for the flux limit $r<19.8$, sample $\mathcal{B}1$ is more stellar mass complete than $\mathcal{A}1$.
Additionally, it is evident from the stellar mass histograms in Fig.~\ref{fig:data_b1_b2} that sample $\mathcal{B}1$ is more stellar mass complete than $\mathcal{B}2$ for the same stellar mass limit $\log \, (M_{\star}/\mathrm{M_{\sun}})^\mathrm{min} = 10.4$. 

This suggests that the different behaviour of MCFs we observe in samples $\mathcal{A}1$ and $\mathcal{B}2$ is due to their stellar mass incompleteness, which in turn comes from the apparent flux limit of the survey and the applied stellar mass limit.

We observe an enhancement in the MCFs of $\mathcal{B}2$ sample relative to $\mathcal{B}1$ at larger scales (Fig.~\ref{fig:result_b1_b2} and right panel of Fig.~\ref{fig:result_b1_b2_ratio}). 
One possible reason could be the influence of the flux limit on the redshift distribution.
In Fig.~\ref{fig:data_b1_b2}, it is clear that the $\mathcal{B}1$ sample is dominated by galaxies at the higher redshift.
But the applied flux limit of $\mathcal{B}2$ reduces the number of galaxies at higher redshift by a factor of almost three.
This effect could have been propagated to the MCFs causing them to deviate from unity at larger scale.

Mass incompleteness effects can also be observed in MCF measurements based on two differently flux-limited SDSS samples (see Fig.~\ref{fig:result_c1_c5}).
The stellar mass MCF of the more incomplete sample $\mathcal{C}5$ shows a lower value than $\mathcal{C}4$ in most of the scales; the sample $\mathcal{C}5$ value varies from $\mathcal{C}4$ on average by $\Delta M_\mathrm{p}(r_\mathrm{p}) = 0.07 \pm 0.06$.
These differences occur even though both samples have the same stellar mass limit.

We do not observe this mass incompleteness effect in GAMA samples, where amplitudes of MCFs do not change significantly with the apparent flux limit.
However, although within error bars, the $\mathcal{C}3$ stellar mass MCF shows a consistently lower value than $\mathcal{C}1$ and $\mathcal{C}2$.
This behaviour is similar to the $\mathcal{C}4$ and $\mathcal{C}5$ stellar mass MCFs.
This can be associated with the variation in the number of galaxies that enter the flux-limited sample even with the same stellar mass limit.
The SDSS sample with the brighter flux limit $r<16.8$ ($\mathcal{C}5$) has 11426 fewer galaxies than that with the fainter limit $r<17.8$ ($\mathcal{C}4$).
At the same time, the differences between the number of galaxies in the GAMA samples ($\mathcal{C}1$, $\mathcal{C}2$, and $\mathcal{C}3$) are very small (see Table~\ref{table:subsamples_definitions}).
Therefore the lack of flux limit dependence of CFs in the GAMA survey can be the result of little variation in the number of galaxies between the GAMA samples $\mathcal{C}1$, $\mathcal{C}2$, and $\mathcal{C}3$. 

To study this effect further, we would have to understand the clustering properties of the galaxies missing in brighter flux-limited samples.
Such studies were previously done by \cite{meneux2008, meneux2009, marulli2013}; however, these studies were only based on measurements of projected CFs.
To understand detailed connections between these galaxies, we would like to measure the behaviour of different MCFs.
As for now, the real data GAMA samples for this kind of studies would only consist of 300-400 galaxies (which are missing in the brighter samples).
So this could only be done using catalogues built from simulations coupled with semi-analytical galaxy formation models.
This kind of study is beyond the scope of the present paper.

\section{Summary and conclusions}\label{sec:conclusion}

In this paper, we studied the environmental dependence of galaxy properties such as luminosity (in $u, g, r, J,$ and $K$ bands), stellar mass, SFR, and sSFR in the redshift range $ 0.1 < z < 0.16$ using a spectroscopic sample of galaxies from the GAMA survey.
We checked which of these properties is a better tracer of the environment, and showed how the results of clustering measurements can be influenced by selecting samples using different properties.
To achieve these aims, we measured the projected correlation function and MCFs in a nearly stellar-mass-complete sample with a flux limit $r < 19.8$ and stellar mass cut $\log \, (M_{\star}/\mathrm{M_{\sun}}) > 9.3$. 
The MCFs were measured using different marks: luminosities in the $u, g, r, J,$ and $K$ bands, stellar mass, SFR, and sSFR.
Additionally we studied the dependence of MCF on the survey flux limit, by repeating the same measurements in two samples with different flux limits ($r < 19.8$ and $r < 17.8$) but the same stellar mass cut ($\log \, (M_{\star}/\mathrm{M_{\sun}}) > 10.4$). 
We also did measurements in samples with GAMA galaxies having mass $\log \, (M_{\star}/\mathrm{M_{\sun}}) > 10.8$ and compared with SDSS galaxies with the same stellar mass cut, but different flux limits.

The summary of our main results and conclusions of our study are as follows:
\begin{itemize}
    \item We observed that different galaxy properties trace the environment differently in the separation scales $r_\mathrm{p} < 10 \, h^{-1} \mathrm{Mpc}$.
    Based on the behaviour of MCFs in Fig.~\ref{fig:result_a1}, we concluded that the close pairs of galaxies are more luminous in the $g, r, J,$ and $K$ bands than distant pairs.
    It is also more probable to find close pairs of massive galaxies (with masses above sample average) than pairs including one less massive galaxy.
    However, this trend is reversed if the luminosity is measured in the $u$ band. 
    The $u$-band luminous galaxies tend to occupy less dense regions and the faint $u$-band galaxies tend to exist in denser regions.
    The same is true for actively star-forming galaxies, which tend to occupy less dense environments.
    \item From the comparisons of the amplitudes of rank-ordered luminosity, stellar mass, SFR, and sSFR MCFs in Fig.~\ref{fig:result_a1}, we concluded that stellar mass is more reliable to trace galaxy environment than the other properties.
    \item We showed that a sample complete in $K$-band luminosity can be a good substitute for a stellar-mass-complete sample.
    But in such a case we tend to miss closer pairs of evolved, red galaxies.
    \item From the similarity in behaviour of different MCFs, we suggested the usefulness of $u$-band luminosity as a proxy of SFR in the context of galaxy clustering.
    \item From the comparative study of MCFs in different samples with the same mass selection, but different apparent magnitude limits, we concluded that closer pairs of star-forming galaxies drop out of the sample when the survey gets shallower in terms of limiting magnitude. 
\end{itemize}

Our measurements are the first of this kind in the redshift range $0.1 < z < 0.16$ and with galaxies as faint as $r < 19.8$.
These measurements can be a useful reference for clustering studies with flux-limited surveys, specially the next generation galaxy surveys such as Vera C. Rubin Observatory \citep[][]{lsst_2009} and Euclid \citep{laureijs2011}.
We intend to extend our measurements using galaxy catalogues from simulations and high-redshift surveys, with which we hope to provide better constraints on models of galaxy formation and evolution.

\begin{acknowledgements}
    %referee
    We thank the anonymous referee for the useful comments and suggestions.
        %grants
        U.S. and A.D. are supported by the Polish National Science Centre grant UMO-2015/17/D/ST9/02121.
        U.S. is supported by Jagiellonian University DSC grant 2019-N17/MNS/000045.
        U.S. and A.P. are supported by the Polish National Science Centre grant UMO-2018/30/M/ST9/00757.
        MB is supported by the Polish National Science Center through grants no. 2020/38/E/ST9/00395, 2018/30/E/ST9/00698, and 2018/31/G/ST9/03388.
        This work is supported by  Polish Ministry of Science and Higher Education grant DIR/WK/2018/12.
        A.H.W is supported by an European Research Council Consolidator Grant (No. 770935).
        %gama survey
        GAMA is a joint European-Australasian project based around a spectroscopic campaign using the Anglo-Australian Telescope. The GAMA input catalogue is based on data taken from the Sloan Digital Sky Survey and the UKIRT Infrared Deep Sky Survey. Complementary imaging of the GAMA regions is being obtained by a number of independent survey programmes including GALEX MIS, VST KiDS, VISTA VIKING, WISE, Herschel-ATLAS, GMRT, and ASKAP providing UV to radio coverage. GAMA is funded by the STFC (UK), the ARC (Australia), the AAO, and the participating institutions. The GAMA website is \url{http://www.gama-survey.org/}.
        Based on observations made with ESO Telescopes at the La Silla Paranal Observatory under programme ID 179.A-2004.
        %tools
        During this research, we made use of Tool for OPerations on Catalogues And Tables \citep[TOPCAT;][]{taylor2005_topcat} and NASA’s Astrophysics Data System Bibliographic Services.
        %computational resources
        This research was supported in part by PLGrid Infrastructure and OAUJ cluster computing facility.
\end{acknowledgements}

\bibliographystyle{aa} 
\bibliography{references} 

\end{document}